\documentclass[iop]{emulateapj}
\newcommand\ush{u_{\rm sh}}
\newcommand\Rx{R_{x1}}

\newcommand\Bpeak{B_{\rm3D, peak}}
\newcommand\xone{x_1}
\newcommand\xtwo{x_2}
\newcommand\cs{c_s}
\newcommand\vA{v_{\rm A}}

\newcommand\Msun{\; {\rm M}_{\odot}}
\newcommand\kms{\; {\rm km}\;{\rm s}^{-1}}
\newcommand\pc{\;{\rm pc}}
\newcommand\kpc{\;{\rm kpc}}
\newcommand\freq{\kms\kpc^{-1}}

\newcommand\yr{\; {\rm yr}}
\newcommand\Myr{\;{\rm Myr}}

\newcommand\Mdotunit{\;\Msun \yr^{-1}}

\newcommand\Surf{\Msun\;{\rm pc^{-2}}}

\newcommand\visunit{\kpc^2\Myr^{-1}}
\newcommand\mG{\mu\rm G}
\newcommand\MBH{M_{\rm BH}}
\newcommand\Min{M_{\rm in}}
\newcommand\Max{M_{R,\phi}}
\newcommand\Bini{B_{\rm3D}}
\newcommand\PhiB{\Phi_{B}}
\newcommand\Rmax{R_{\rm max}}
\newcommand\Rmin{R_{\rm min}}
\newcommand\RCR{R_{\rm CR}}
\newcommand\ILR{R_{\rm ILR}}
\newcommand\IILR{R_{\rm IILR}}
\newcommand\OILR{R_{\rm OILR}}
\newcommand\Omb{\Omega_{b}}
\newcommand\vphi{v_\phi}
\newcommand\ip{i_p}
\newcommand\Sigfl{\Sigma_{\rm floor}}
\newcommand\rhobar{\rho_{\rm bar}}
\newcommand\rhobul{\rho_{\rm bul}}
\newcommand\uperp{u_\bot}
\newcommand\upara{u_\|}
\newcommand\Mperp{{\mathcal M}_\bot}

\newcommand\aSig{\langle\Sigma\rangle}
\newcommand\aBth{\langle B_{\rm 3D}\rangle}
\newcommand\bSig{\Sigma_{\rm avg}}
\newcommand\pSig{\Sigma_{\rm peak}}
\newcommand\Rring{R_{\rm ring}}

\newcommand\Mdot{\dot{M}}

\newcommand\simgt{\lower.5ex\hbox{$\; \buildrel > \over \sim \;$}}
\newcommand\simlt{\lower.5ex\hbox{$\; \buildrel < \over \sim \;$}}

\shorttitle{Magnetized Barred Galaxies}
\shortauthors{Kim et al.}
\begin{document}

\title{Two-Dimensional Magnetohydrodynamic Simulations of Barred Galaxies}
\author{Woong-Tae Kim\altaffilmark{1,2,3,4} \&
James M.\ Stone\altaffilmark{4}}%%
\affil{$^1$Center for the Exploration of the Origin of the Universe
(CEOU), Astronomy Program, Department of Physics \& Astronomy,
\\ Seoul National University, Seoul 151-742, Republic of Korea}%%
\affil{$^2$FPRD, Department of Physics
\& Astronomy, Seoul National University, Seoul 151-742, Republic of
Korea}%%
\affil{$^3$Institute for Advanced Study, Einstein Drive, Princeton,
NJ 08540, USA}%%
\affil{$^4$Department of Astrophysical Sciences, Princeton
University, Princeton, NJ 08544, USA}%%
\email{wkim@astro.snu.ac.kr,  jstone@astro.princeton.edu}%%
\slugcomment{Accepted for publication in the ApJ}

\begin{abstract}
Barred galaxies are known to possess magnetic fields that may affect
the properties of bar substructures such as dust lanes and nuclear
rings. We use two-dimensional high-resolution magnetohydrodynamic
(MHD) simulations to investigate the effects of magnetic fields on
the formation and evolution of such substructures as well as on the
mass inflow rates to the galaxy center. The gaseous medium is
assumed to be infinitesimally-thin, isothermal,
non-self-gravitating, and threaded by initially uniform, azimuthal
magnetic fields.  We find that there exists an outermost
$\xone$-orbit relative to which gaseous responses to an imposed
stellar bar potential are completely different between inside and
outside.  Inside this orbit, gas is shocked into dust lanes and
infalls to form a nuclear ring. Magnetic fields are compressed in
dust lanes, reducing their peak density. Magnetic stress removes
further angular momentum of the gas at the shocks, temporarily
causing the dust lanes to bend into an ``L'' shape and eventually
leading to a smaller and more centrally distributed ring than in
unmagnetized models. The mass inflow rates in magnetized models
correspondingly become larger, by more than two orders of magnitude
when the initial fields have an equipartition value with thermal
energy, than in the unmagnetized counterparts. Outside the outermost
$\xone$-orbit, on the other hand, an MHD dynamo due to the combined
action of the bar potential and background shear operates near the
corotation and bar-end regions, efficiently amplifying magnetic
fields.   The amplified fields shape into trailing magnetic arms
with strong fields and low density. The base of the magnetic arms
has a thin layer in which magnetic fields with opposite polarity
reconnect via a tearing-mode instability. This produces numerous
magnetic islands with large density which propagate along the arms
to turn the outer disk into a highly chaotic state.
\end{abstract}

\section{Introduction\label{sec:intro}}

One of the characteristic features of barred galaxies is the
existence of gaseous substructures such as a pair of dust lanes, a
nuclear ring, and nuclear spirals, visible in optical images and
radio maps (e.g., \citealt{pea17,san61,san76,but96,
mar03a,mar03b,pri05,mar06,com09, hsi11}).  Dust lanes located at the
leading side of the bar are interpreted as shocks in the gas flows,
as evidenced by the enhanced radio emissions and sharp velocity
jumps across them (see, e.g., \citealt{but96} and references
therein). The curvature of dust lanes depends on the bar strength
and aspect ratio, while largely independent of its pattern speed and
galaxy mass \citep{kna02,com09}. Nuclear rings are the regions of
high gas density and usually populated with numerous \ion{H}{2}
regions, indicative of recent starburst activities (e.g.,
\citealt{but86,hel94,mao01,maz08,hsi11}). They have a size typically
of $1\kpc$, but appear to be larger and wider in galaxies with less
compact mass distribution \citep{maz11}. Nuclear spirals located
inside the rings are believed to be a channel for gas inflows to the
galaxy center, potentially powering active galactic nuclei (AGN)
(e.g., \citealt{shl90,reg99,kna00,van10}).

Since the gas flows shaping bar substructures are intrinsically
nonlinear, numerical simulations have been a powerful tool to study
their formation, evolution, and associated dynamics (e.g.,
\citealt{san74,ath92b,pin95,eng97,pat00,reg03,reg04,mac04,ann05,tha09}).
In particular, \citet{ath92b} showed that a non-axisymmetric bar
potential induces dust-lane shocks at downstream from the bar major
axis and that the shocks become more straight under a stronger bar.
Dust lanes tend to become shorter and located closer to the bar
major axis when the gas sound speed is larger \citep{eng97,pat00}.
The shocked gas at dust lanes loses angular momentum and flows
inward to form a nuclear ring at the position where the external
gravity is balanced by the centrifugal forces
\citep{shl90,pin95,reg03}. Using a smoothed particle hydrodynamics
method, \citet{ann05} and \citet{tha09} showed that the formation of
strong nuclear spirals supported by shocks requires a large sound
speed and the presence of a central black hole (BH), although their
models were unable to resolve weak spirals in the nuclear regions
due to a small number of particles (see also \citealt{mac04}).

Very recently, \citet[][hereafter Paper I]{kim12} ran
high-resolution simulations using the grid-based CMHOG code on a
cylindrical geometry.  Paper I found that the original CMHOG code
used by \citet{pin95} contained errors in the evaluation of the bar
forces, compromising their numerical results  as well as others
(e.g., \citealt{reg03,mac04}) who unknowingly adopted the original
code with the errors. Using a corrected version of the code, Paper I
found that when the gas sound speed $\cs$ is small, nuclear rings
are narrow and decoupled from nuclear spirals, making the mass
inflow rates $\Mdot$ toward the galaxy center extremely small ($\sim
10^{-4} \Mdotunit$). On the other hand, strong thermal perturbations
in models with high $\cs$ make the rings broad, leading to
$\Mdot\simgt 10^{-2}\Mdotunit$. Paper I also found that the shape
and strength of nuclear spirals depend sensitively on the gas sound
speed and the BH mass $\MBH$ such that they are leading if
both $\cs$ and $\MBH$ are small, weak trailing if $\cs$ is small and
$\MBH$ is large, and strong trailing if both $\cs$ and $\MBH$ are
large: nuclear spirals are readily destroyed by the ring material on
eccentric orbits when $\cs$ is large and $\MBH$ is small.

While aforementioned work improved our understanding on gas dynamics
associated with bar substructures, they are without one of the most
important ingredients of the interstellar medium (ISM), namely
magnetic fields that are pervasive in disk galaxies (e.g.,
\citealt{bec96,bec09}). By observing barred galaxies in radio
polarized emissions, \citet{bec05} found that magnetic fields are
strong in dust lanes and nuclear rings, suggesting that they are
dynamically important in shaping bar substructures (see also
\citealt{bec99,bec02}).  Some barred galaxies such as NGC 1365, NGC
1097, NGC 1365 are observed to possess ``magnetic arms''
characterized by stronger fields and lower density than surrounding
regions \citep{bec02,bec05}. In addition, magnetic fields exert
torque that is much larger than gravitational torque, providing an
efficient means to transport gas from a nuclear ring inward to feed
an AGN \citep{bec05}. Therefore, the effect of magnetic fields on
structure and evolution of barred galaxies should not be ignored.

There have been a number of numerical studies on magnetic field
distributions in barred galaxies.  Depending on how magnetic fields
are treated, they can be categorized into two groups: (1) those
based on mean-field dynamo theories  (e.g.,
\citealt{otm97,otm02,mos_etal98,mos99,mos01,mos07}) and (2) those
using full magnetohydrodynamic (MHD) simulations
\citep{kul09,kul10,kul11}. Although numerical models from the first
group are successful in obtaining synthetic polarization maps and
overall field morphologies comparable to observations, they rely on
parameterized turbulent terms in the induction equation that are
uncertain. More importantly, the mean-field dynamo models take
velocity fields from hydrodynamic or $N$-body sticky-particle
simulations and evolve magnetic fields passively, without
considering the back reaction of magnetic fields to the gas. On the
other hand, MHD models in the second group naturally handle gas
responses to the embedded magnetic fields.  For instance,
\citet{kul10} found that dynamical effects of magnetic fields lead
to the formation of low-density magnetic arms in the regions outside
the bar, which is clearly unseen in the mean-field dynamo models
(see also \citealt{kul09,kul11}).  Even without the prescription for
the mean-field dynamo,  magnetic energy in these MHD models grows
due to shear and compression by a factor of about 15 relative to the
initial value over a 1 Gyr period \citep{kul09}.

Because the MHD models mentioned above focused on magnetic
structures in the outer regions, however, the effects of magnetic
fields on bar substructures and mass inflow rates have yet to be
explored. Can magnetic fields make dust lanes stronger? Do they make
nuclear rings smaller or larger?  Do nuclear spirals survive in the
presence of magnetic fields? How do the mass inflow rates change as
the field strength varies?   In order to address these questions, we
in this paper run high-resolution MHD simulations using the Athena
code \citep{gar05,sto08,sto09}. This work is a straightforward
extension of Paper I by including magnetic fields. Our models are
two dimensional, assuming a razor-thin disk. This allows to better
resolve the in-plane direction than in the three-dimensional models
considered by \citet{kul10}.  On the other hand, our two-dimensional
models are unable to capture the dynamical consequences of Parker
instability and other processes that involve the vertical direction.
In addition to studying the magnetic effects on substructures in the
bar regions, we identify an MHD dynamo mechanism that amplifies
magnetic fields, eventually leading to magnetic arms in the outer
regions.

This paper is organized as follows. In Section 2, we describe our
numerical methods and model parameters, and present numerical
resistivity of the Athena code. In Section 3, we make a comparison
between the results of the Athena and CMHOG codes on hydrodynamic
models with identical parameters. In Section 4, we present the
results of MHD simulations on the bar regions, focusing on the
effects of magnetic fields on bar substructures and mass inflow
rates. In Section 5, we study evolution of gas and magnetic fields
in the outer regions.  In Section 6, we conclude with a summary and
discussion of our numerical results.

\section{Models and Methods}\label{sec:model}

We consider a magnetized gaseous disk and study its responses to an
imposed non-axisymmetric bar potential. Gas dynamics without the
effect of magnetic fields was presented in Paper I. As in Paper I,
the disk is initially uniform, isothermal, vertically-thin, and
non-self-gravitating. The bar is assumed to rotate rigidly about the
galaxy center with a fixed pattern speed
$\mathbf{\Omb}=\Omb\mathbf{\hat z}$. We solve the dynamical
equations in the frame corotating with the bar in the $z=0$ plane.
The equations of magnetohydrodynamics in the rotating frame
integrated along the vertical direction are
\begin{equation}\label{eq:con}
\left(\frac{\partial}{\partial t}  + \mathbf{u}\cdot\nabla \right) \Sigma
= - \Sigma\nabla\cdot\mathbf{u},
\end{equation}
%\begin{equation}\label{eq:mom}
\begin{mathletters}\label{eq:mom}
\begin{eqnarray}
\left(\frac{\partial}{\partial t}  + \mathbf{u}\cdot\nabla \right) \mathbf{u}
& =&  -\frac{\cs^2}{\Sigma} \nabla \Sigma
 +  \frac{1}{4\pi\Sigma} (\nabla\times \mathbf{B}) \times \mathbf{B}
\nonumber \\
& -& \nabla \Phi_{\rm ext} + \Omb^2
\mathbf{R} - 2\mathbf{\Omb}\times \mathbf{u},
\end{eqnarray}
\end{mathletters}
%\end{equation}
\begin{equation}\label{eq:ind}
\frac{\partial \mathbf{B}}{\partial t} = \nabla\times
(\mathbf{u} \times \mathbf{B}).
\end{equation}
Here, $\Sigma$, $\mathbf{u}$, and $\cs$ denote the surface density,
velocity in the rotating frame, and the sound speed in the gas,
while $\mathbf{B}$ represents the midplane value of the
three-dimensional magnetic field times the square root of the ratio
of surface density to midplane volume density (e.g.,
\citealt{kim01}). Note that the induction equation (\ref{eq:ind}) in
the rotating frame has the same form as in the inertial frame (e.g.,
\citealt{kol02}). The velocity $\mathbf{v}$ in the inertial frame is
obtained from $\mathbf{v} = \mathbf{u} + R\Omb\mathbf{\hat \phi}$.
We do not consider star formation and the associated gas recycling
in the present work.

The external gravitational potential, represented by $\Phi_{\rm
ext}$ in equation (\ref{eq:mom}), and the associated rotation curve
are taken identical to those in Paper I.  Here we describe these
briefly. The external potential consists of four components: a
stellar disk modeled by a Kuzmin-Toomre disk, a central bulge
modeled by a modified Hubble profile with central density $\rhobul$,
a non-axisymmetric bar modeled by a \citet{fer87} spheroid with
central density $\rhobar$, and a central supermassive BH with mass
$\MBH$. Without a BH, the rotation velocity $v_c$ rises steeply with
the galactocentric radius $R \simlt 1\kpc$ and attains a
more-or-less constant value of $\sim200\kms$ at $R\simgt 6\kpc$. The
presence of a BH affects $v_c$ near the central parts, resulting in
$v_c\propto (\MBH/R)^{1/2}$ at $R\simlt 0.1\kpc$. The bar has a
patten speed of $\Omb=33\freq$, which places the corotation
resonance (CR) at $\RCR=6\kpc$. In order to avoid abrupt responses
of the gas due to a sudden introduction of the bar, we increase
$\rhobar$ linearly with time over one bar revolution time of
$2\pi/\Omb=186$ Myr, while keeping the net density $\rhobar+\rhobul$
fixed.  This ensures that shape of $\Phi_{\rm ext}$ averaged along
the azimuthal direction is not changed much with time.

Our model disks initially have uniform surface density
$\Sigma_0=10\Surf$ and isothermal sound speed of $\cs=5\kms$.
Magnetic fields are initially in plane, uniform, and purely
azimuthal. The field strength is parameterized by the dimensionless
plasma parameter
\begin{equation}\label{eq:beta}
\beta \equiv \frac{8\pi \cs^2\Sigma}{B^2}=\frac{8\pi \cs^2\rho}{\Bini^2}.
\end{equation}
where $\rho$ and $\Bini$ are the midplane values of volume density
and three-dimensional magnetic field strength.
In terms of dimensional units
\begin{equation}\label{eq:bini}
\Bini = \frac{4.6\mG}{\sqrt{\beta}}
\left(\frac{\cs}{5\kms}\right)
\left(\frac{\Sigma}{10\Surf}\right)^{1/2}
\left(\frac{H}{100\pc}\right)^{-1/2},
\end{equation}
where $H=\Sigma/(2\rho)$ is the scale height of the gaseous disk.
Note that the initial disk with constant $\beta_0$ has a mass-to-flux
ratio $M/\PhiB = (\pi\beta_0\Sigma_0/8\cs^2)^{1/2}R$ that is
increasing linearly with $R$.

In order to focus on the effects of magnetic field strength and
rotation curve at the central parts, we consider eight models with
differing $\MBH$ and $\beta_0$, as listed in Table \ref{tbl:model}.
The strength of magnetic fields is chosen to vary from
sub-equipartition ($\beta_0=10$) to equipartition ($\beta_0=1$)
values with thermal energy. We also consider hydrodynamic models
with $\beta_0=\infty$ to make comparisons with the results of Paper
I. Models with a prefix bh7 possess a central BH with $\MBH=4\times
10^7\Msun$, with the $\Omega-\kappa/2$ curve attaining the local
maximum and minimum at  $\Rmax=0.53\kpc$ and $\Rmin=0.19\kpc$,
respectively (see Paper I). The bh7 models have only a single inner
Lindblad resonance (ILR) at $\ILR\approx2\kpc$ where
$\Omega-\kappa/2=\Omb$. On the other hand, bh0 models with no BH
have two ILRs at $\IILR =0.19\kpc$ and $\OILR\approx 2\kpc$, and the
$\Omega-\kappa/2$ curve peaks at $\Rmax=0.53\kpc$. We take Model
bh7MHD01 with $\MBH=4\times 10^7\Msun$ and $\beta_0=1$ as our
standard model.

%table1
\begin{deluxetable}{lcc}
\tabletypesize{\footnotesize}
\tablewidth{0pt}
\tablecaption{Model Parameters\label{tbl:model}}
\tablehead{
\colhead{Model} &
\colhead{$\MBH (\Msun)$} &
\colhead{$\beta_0$}
}
\startdata
bh7MHD01 & $4\times10^7$  & 1  \\
bh7MHD03 & $4\times10^7$  & 3  \\
bh7MHD10 & $4\times10^7$  & 10  \\
bh7HD    & $4\times10^7$  & $\infty$  \\
\hline
bh0MHD01 & 0 & 1  \\
bh0MHD03 & 0 &  3  \\
bh0MHD10 & 0 &  10  \\
bh0HD    & 0 & $\infty$
\enddata
\end{deluxetable}

We solve the time-dependent ideal MHD equations
(\ref{eq:con})-(\ref{eq:ind}) using a modified version of the Athena
code in Cartesian coordinates \citep{gar05,sto08,sto09}. Athena
utilizes a higher order Godunov scheme which conserves mass,
momentum, and magnetic flux within machine precisions. It also
provides several different schemes for integration in time, spatial
reconstruction, and solution of the Riemann problem. We take the van
Leer algorithm with piecewise linear reconstruction and first order
flux correction.  For a Riemann solver, we use the exact nonlinear
solver for hydrodynamic models and an approximate HLLD-type
nonlinear solver for MHD runs.

%fig1
\begin{figure*}
\epsscale{1} \plotone{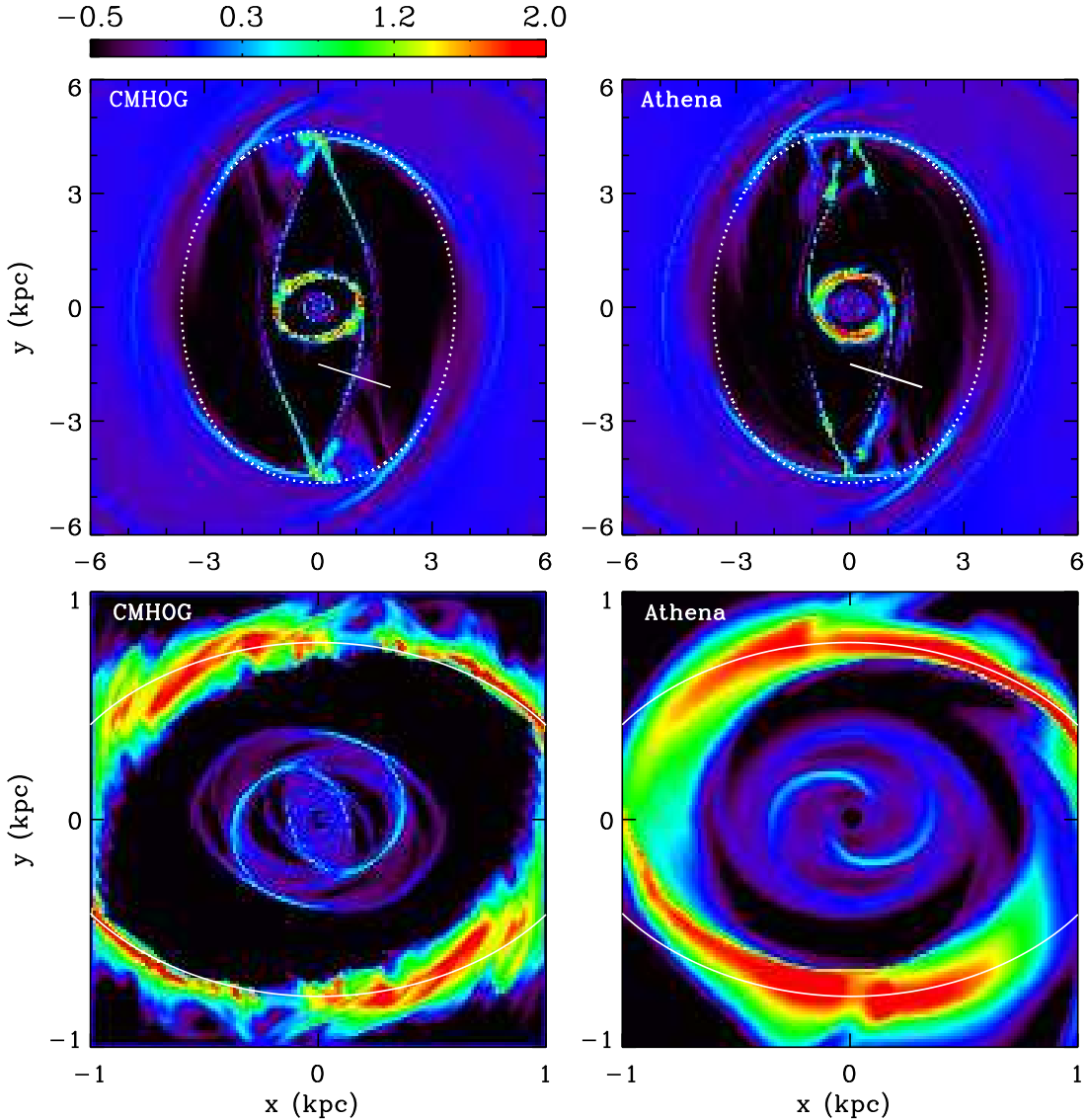} \caption{Comparative snapshots of
surface density in logarithmic scale from hydrodynamic runs with no
BH using CMHOG (left) and Athena (right) at $t=300$ Myr.  The upper
panels cover the corotation resonance, while the lower panels zoom
in the central $1\kpc$ regions.  The dotted ovals in the upper
panels draw the outermost $\xone$-orbit crossing the $x$- and
$y$-axes at $x_c=3.6\kpc$ and $y_c=4.7\kpc$, respectively, while the
short line segments indicate a slit perpendicular to the dust lane,
along which density and velocity are measured in Figure
\ref{fig:compprof}a,b. The solid curves in the lower panels draw the
$\xtwo$-orbit with $x_c=1.2\kpc$. \label{fig:compsnap}}
\end{figure*}
Our simulation domain is a square box with side $L=30\kpc$ in each
direction. We set up a uniform Cartesian grid with $4096\times4096$
zones over $|x|, |y|\leq L/2$.  The corresponding grid spacing is
$\Delta x=\Delta y=7.3\pc$, about an order magnitude smaller than
models in \citet{kul10}.  These high-resolution runs are necessary
to explore properties of bar substructures in detail. We apply the
outflow boundary conditions at the domain boundaries (i.e., at
$|x|=L/2$ or $|y|=L/2$). To measure the mass inflow rates toward the
galaxy center, we place a central hole with radius of 20 pc and
define a density floor $\Sigfl$ inside the hole. At each time, we
calculate the gas mass $\Min$ flown into the hole by integrating the
excess interior density as $\Min = \int_{\rm hole} (\Sigma-\Sigfl)
dxdy$, and then reset the interior density to $\Sigfl$. The
mass inflow rate is then obtained from $\Mdot=\Min/\Delta t$, where
$\Delta t$ is the computational timestep. In this paper, we present
the results with $\Sigfl= 10^{-4}\Surf$, but we checked that $\Mdot$
is insensitive to the choice of $\Sigfl$ as long as it is taken
sufficiently small.

The inflowing gas may carry magnetic fields to the galaxy center, as
well. In order to allow for $B$-field accretion, we let the magnetic
fields vanish inside the central hole while preserving the
divergence-free condition. This is achieved by calculating the
vector potential at each time, cutting it off to a constant value
inside the hole, and then recalculating the $\mathbf{B}$ vector from
the modified vector potential. We note that taking $B=0$ arbitrarily
inside the hole sometimes makes $B$-fields bend abruptly near the
hole boundary. In addition, density in the vicinity of the hole
frequently becomes very small, greatly limiting the computational
timestep by increasing the Alfv\'en speed.  In order to minimize
these spurious effects, we utilize the non-ideal Ohmic dissipation term
included in Athena by taking magnetic resistivity $\eta=1\visunit$
in the circular region at $R\leq 40\pc$, while the rest
of the simulation domain is kept non-resistive. This results in a
gradual dissipation of $B$-fields in the region surrounding the
hole, allowing us to run simulations for a long period of time.

Finally, we remark on the numerical resistivity of the Athena code
that we use.  It is difficult to estimate the magnetic diffusivity
of the real ISM accurately since it is very clumpy and diffusivity
changes a lot between the different phases with different
temperature and different degree of ionization. If turbulence
dominates the diffusion and dissipation of the magnetic fields, it
is perhaps of the order of $\eta_T\sim10^{-4}-10^{-3}\visunit$
(e.g., \citealt{par71,cam94,kul09}).  The associated magnetic
Reynolds number is $R_m= v_cL/\eta_T \sim 10^4-10^5$, much larger
than unity, so that taking the $R_m\rightarrow\infty$ limit, as we
do in the present work except for the small central regions, is a
useful starting point. Nevertheless, we note that there is non-zero
(but small) numerical diffusivity over the entire simulation domain
introduced by the numerical scheme. In Appendix, we use
linear-amplitude magnetosonic waves to evaluate the coefficients of
numerical viscosity as well as magnetic diffusivity of the Athena
code. For $\Delta x=7.3\pc$ and the wavelength $\lambda\sim0.5\kpc$
typical for the tearing-modes of magnetic reconnection occurring in
the outer regions (see Section \ref{sec:outer}), the numerical
resistivity is estimated to be $\eta_n \sim 6\times10^{-7}\visunit$,
about three orders of magnitude smaller than $\eta_T$.  The
associated magnetic diffusion time is $\tau_{\rm mag}=\Delta
x^2/\eta_n \sim 100\Myr$, indicating that the non-ideal effect of
numerical resistivity on the $B$-field evolution can be important if
the field strength changes considerably over the grid scale.

\section{Comparison between CMHOG and Athena Results}\label{sec:comp}

As mentioned above, Paper I explored hydrodynamical models of
barred galaxies using the corrected version of the cylindrical CMHOG
code that originally contained a bug in calculating the bar forces
\citep{pin95}.  On the other hand, the present calculations make use
of the Athena code on a uniform Cartesian grid. Since the current
models include two hydrodynamic models with the same parameters as
in Paper I, it is interesting to compare the results from two
different codes.  This will check not only the reliability of the
Athena runs but also the validity of the
corrected bar forces in Paper I. Note that by employing a
non-uniform grid, the CMHOG models in Paper I has a higher (lower)
spatial resolution at $R<1.1\kpc$ ($R>1.1\kpc$) than the current
Athena runs with a uniform grid spacing.

Overall evolution of hydrodynamic models is described in Paper I:
here we summarize the main features. As the bar potential is slowly
introduced, initially-circular gas orbits are perturbed, forming a
pair of overdense ridges at the downstream side of the bar major
axis.  As the bar potential grows further, the overdense ridges
eventually develop into shocks, appearing as dark dust lanes in
optical images of barred galaxies. Gas loses a significant amount of
angular momentum at the shocks and thus moves radially in, forming a
nuclear ring at the position where the centrifugal force balances
the external gravity. In addition, these shocks are curved and thus
able to generate vorticity. Vorticity grows secularly with time by
the successive passages across shocks. When it achieves substantial
amplitudes, it causes the shock fronts to wiggle, producing small
clumps with high vorticity.  The whole process of clump formation is
analogous to the wiggle instability of spiral shocks identified by
\citet{wad04} (see also \citealt{kim06}). These clumps are added to
the nuclear ring. As clumps collide with each other, the ring
becomes homogeneous gradually.

Figure \ref{fig:compsnap} compares the snapshots of Model bh0HD in
the present work with those of Model cs05bh0 from Paper I at $t=300$
Myr when the bar substructures reach a quasi-steady state. The bar
is oriented vertically along the $y$-axis, and the gas inside
(outside) the CR at $R=6\kpc$ is rotating in the counterclockwise
(clockwise) direction.  The images in the upper panels cover the
whole region inside the corotation, while the lower panels enlarge
the central $1\kpc$ regions. In the upper panels, dotted curves draw
the outermost $\xone$-orbit that cuts the $x$- and $y$-axes at
$x_c=3.6$ and $y_c=4.7\kpc$, respectively, and has a Jacobi energy
of $E_J=-1.24\times 10^5\;{\rm(km\;s^{-1})^2}$ under our adopted
external potential (e.g., \citealt{ath92a}). Inside this orbit,
there are families of closed $\xone$- and $\xtwo$-orbits that cross
each other, resulting in shocks at the leading side of the bar major
axis. As gas loses angular momentum at the shocks and moves toward
the central parts, the bar regions becomes gradually evacuated.
Outside the outermost $\xone$-orbit, no closed orbit exists: gas
there simply moves on slightly perturbed, near-circular orbits,
without inducing shocks.  The gas moving in slowly due to the bar
torque from outside is subsequently trapped in the outermost
$\xone$-orbit, and stays in it.  Therefore, the outermost
$\xone$-orbit acts as a barrier that inflowing gas cannot penetrate.
Other than weak trailing spirals produced by the bar toque that
emerge from the downstream side of the bar ends, the outer
regions are almost featureless in hydrodynamic models.

%fig2
\begin{figure}
\epsscale{1.2} \plotone{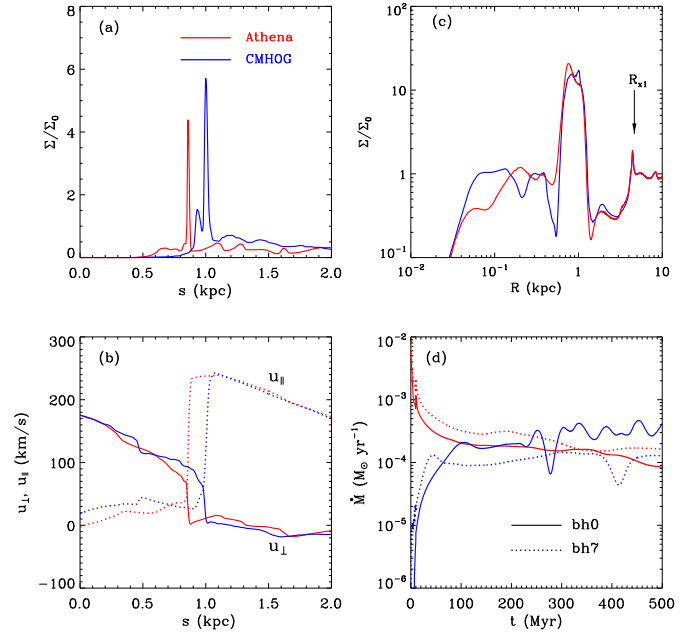} \caption{Comparisons between the
Athena and CMHOG results for the distributions of (a) surface
density and (b) velocity $\uperp$ perpendicular and $\upara$
parallel to the dust-lane shocks along the slit shown in Figure
\ref{fig:compsnap}, (c) for the radial distribution of gas density
averaged both azimuthally and temporally over $t=300-500$ Myr, and
(d) for the temporal variations of the mass inflow rates. In (c),
$\Rx=4.7\kpc$ with an arrow marks the maximum radial extent of the
outermost $\xone$-orbit. \label{fig:compprof}}
\end{figure}

%fig3
\begin{figure*}
\epsscale{1.1} \plotone{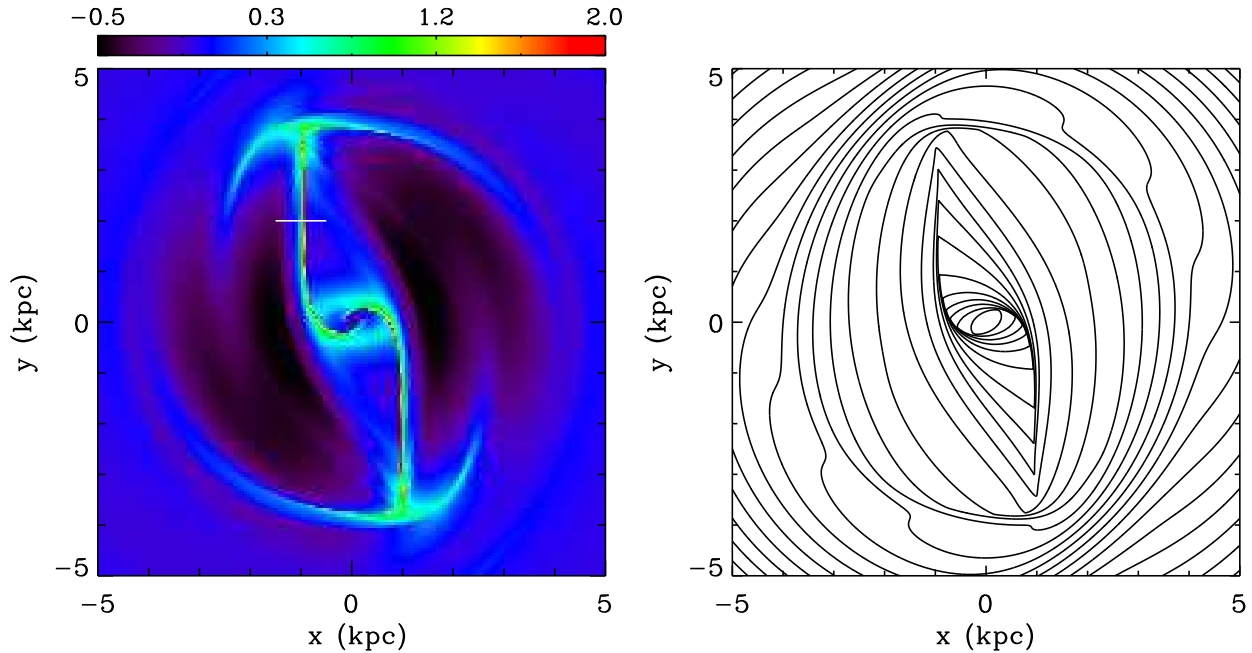} \caption{Logarithm of the density
distribution (left) and magnetic field configurations (right) of
Model bh7MHD01 with $\MBH=4\times10^7\Msun$ and $\beta_0=1$ at
$t=150$ Myr. The shot line segment at $-1.5 \leq x\leq -0.5$ and
$y=2\kpc$ indicates a slit perpendicular to the dust lane, along
which fluid variables are measured in Figure \ref{fig:shock}.
\label{fig:snap_t150}}
\end{figure*}

In the lower panels, solid curves draw an $\xtwo$-orbit with
$x_c=1.2\kpc$ that approximately follows the nuclear ring. The
morphological agreement between the CMHOG and Athena results is
fairly good, although the ring in the CMHOG run is more clumpy and
slightly larger.  This is because the Athena run has a higher
resolution at $R>1.1\kpc$ where dust lanes form.  This makes the
dust-lane shocks in the Athena runs better resolved, leading to
higher angular momentum loss and a smaller ring than in the CMHOG
counterpart. In addition, the wiggle instability of the dust lanes
is more vigorous in the Athena run since it grows faster at smaller
scales. Hence, the Athena run produces more clumps that collide
frequently, resulting in a smoother ring than in the CMHOG run.
Since the Athena run has a lower resolution at $R<1.1\kpc$, on the
other hand, it harbors weaker nuclear spirals than in the CMHOG run,
as Figure \ref{fig:compsnap} shows. The peak density relative to the
mean and the pitch angle of the spirals at $R=0.25\kpc$ are
$\pSig/\bSig=3.7$ and $\ip=-30^\circ$ in the Athena run, which can
be compared with $\pSig/\bSig=7.9$ and $\ip=-34^\circ$ in the CMHOG
run.

Figure \ref{fig:compprof} provides more quantitative comparisons
between the Athena and CMHOG results. Figure \ref{fig:compprof}a,b
plot the distributions of surface density and velocities along the
slit indicated in Figure \ref{fig:compsnap}. The slit starts from
$(x,y)=(0, -1.5\kpc)$ and ends at $(1.9, -2.1)\kpc$. The dust-lane
shocks in Athena are slightly weaker and located $0.13\kpc$ closer
to the bar major axis than in CMHOG.  This is consistent with a
smaller nuclear ring in the Athena run, since the inner ends of the
dust lanes are connected to the nuclear ring. The overall velocity
profiles as well as the amount of the velocity jumps from the Athena
and CMHOG runs agree well with each other. Figure
\ref{fig:compprof}c compares the radial profiles of the mean density
averaged both azimuthally and temporally over $t=300-500$ Myr.
Although the density-weighted ring radius is slightly smaller in the
Athena run, the difference is only $\sim0.05\kpc$, again
demonstrating good agreement between the CMHOG and Athena results.
The black arrow marked by $\Rx=4.7\kpc$ indicates the maximum radial
extent of the outermost $\xone$-orbit, outside of which density is
relatively unperturbed. A small spike of the mean density around
$R=\Rx$ is caused by the trapping of inflowing gas from outside.

Finally,  Figure \ref{fig:compprof}d compares the mass inflow rates
for bh0 and bh7 models as functions of time.  Because of difference
in the way handling fluid variables inside the central hole, the
temporal behavior of $\Mdot$ at early time is quite different
between the Athena and CMHOG results, although it rapidly converges
to a more-or-less constant value. In the CMHOG runs, the bh0 model
has larger $\Mdot$ at $t>300$ Myr than in the bh7 model, while the
former has slightly smaller $\Mdot$ in the Athena runs.
Nevertheless, $\Mdot$ at $t>100$ Myr from the Athena and CMHOG runs
agrees with each other, on average, within a factor of 2, implying
that the current method of measuring $\Mdot$ by placing a central hole
is reliable.

To conclude this section, despite the CMHOG and Athena codes are
quite different in terms of coordinate geometry as well as ways of
updating fluid variables, the results from both codes are in good
agreement. This confirms not only that the revised force
transformation in CMHOG used in Paper I is correct, but also that
dynamics of a rotating disk can be reliably handled by the Cartesian
Athena code that we use.

\section{Bar Regions}\label{sec:inner}

We now turn to the effects of magnetic fields on bar substructures
and mass inflow rates.  Evolution of the gas outside the outermost
$\xone$-orbit is quite distinct from that inside.
In this section we focus on the bar regions
(i.e., inside the outermost $\xone$-orbit):
the changes of magnetic structures in the outer
regions will be presented in Section \ref{sec:outer}.

\subsection{Overall Evolution}

\begin{figure}
\epsscale{1.1} \plotone{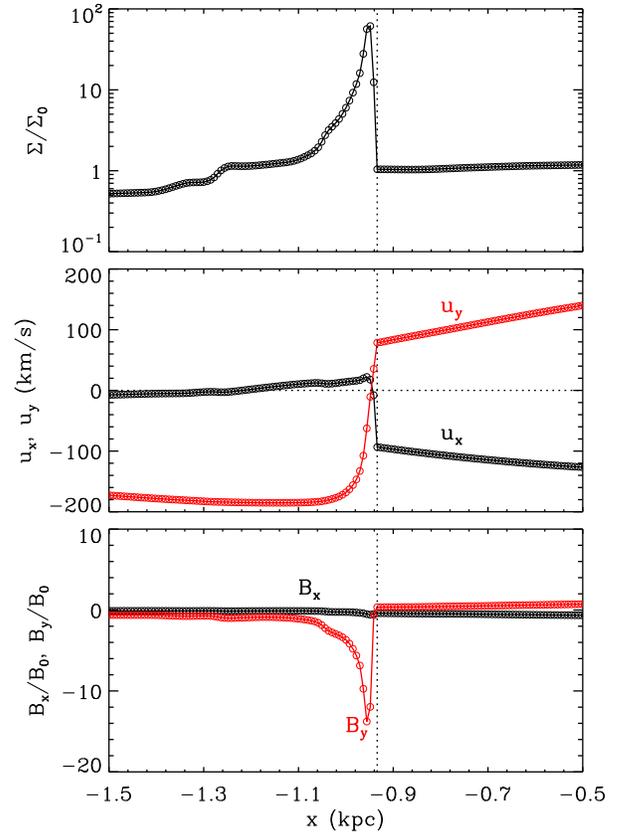} \caption{Profiles of surface
density, velocity, and magnetic fields along the slit shown in
Figure \ref{fig:snap_t150}.  Note the $x$- and $y$-components of the
velocity and magnetic fields are in the direction perpendicular and
parallel to the shocks, respectively. Each circle corresponds to
an individual pixel value. \label{fig:shock}}
\end{figure}

Early evolution in the bar regions of our standard model bh7MHD01
(with $\beta_0=1$ and $\MBH=4\times10^7\Msun$) is not much different
from the hydrodynamic counterpart in that the bar potential induces
crowding of gas orbits to create overdense ridges that develop into
dust-lane shocks. The presence of magnetic fields (1) reduces the
shock density and (2) provides Maxwell stress for the gas at the
shocks, further removing angular momentum from it. To illustrate
this, Figure \ref{fig:snap_t150} plots a surface density snapshot
together with the magnetic field configuration of Model bh7MHD01 at
$t=150$ Myr. Note the magnetic field lines are \emph{approximate}
proxy to instantaneous gas streamlines. Figure \ref{fig:shock} plots
the variations of gas surface density, velocities, and magnetic
fields along the slit at $y=2\kpc$ shown in Figure
\ref{fig:snap_t150}. Each circle corresponds to an individual grid
point. At this time, the shocks are almost parallel to the $y$-axis,
and are moving toward the bar major axis at velocity
$\ush\simeq30\kms$ relative to the bar.

The perpendicular Mach number of the flows along the slit relative
to the shock is $\Mperp=(u_{x,1} -\ush)/\cs\simeq 13$. Here and
hereafter, the subscripts $1$ and $2$ denote the preshock and
postshock value, respectively. With
$\beta_{y,1}=8\pi\cs^2\Sigma_1/B_{y,1}^2=8$, the usual jump
condition for one-dimensional stationary MHD shocks\footnote{ For
perpendicular isothermal shocks, the density compression factor $r$
is a positive root of the quadratic equation $r^2 + (1+\beta_1)r -
\beta_1\Mperp^2=0$ (e.g., \citealt{pri82}).} yields the compression
factor $r=\Sigma_2/\Sigma_1=B_{y,2}/B_{y,1}=32.5$ and
$B_{x,2}/B_{x,1}=1$. Figure \ref{fig:shock} shows that
$\Sigma_2/\Sigma_1\simeq57$ and $B_{y,2}/B_{y,1}\simeq-40$, and
$B_{x,2}/B_{x,1}\simeq1$, indicating that the density compression of
the dust-lane shocks is more than steady one-dimensional shocks
allow. More importantly, the sign of $B_y$ is reversed across the
shock. This is because the shocks are intrinsically two dimensional
in the sense that the streamlines diverge before the shock and bend
abruptly to align themselves parallel to the dust lanes immediate
after the shock. The bending of the streamlines at the shock is
caused by a strong inflow of the gas along the dust lanes from the
bar ends. This postshock inflow rotates the magnetic fields at the
shock, making the sign of $B_y$ reversed. Magnetic tension forces
from the bent $B$-fields are highly efficient to remove angular
momentum from the gas moving across shocks, as we will show below.

%fig4

%fig5
\begin{figure}
\epsscale{1.1} \plotone{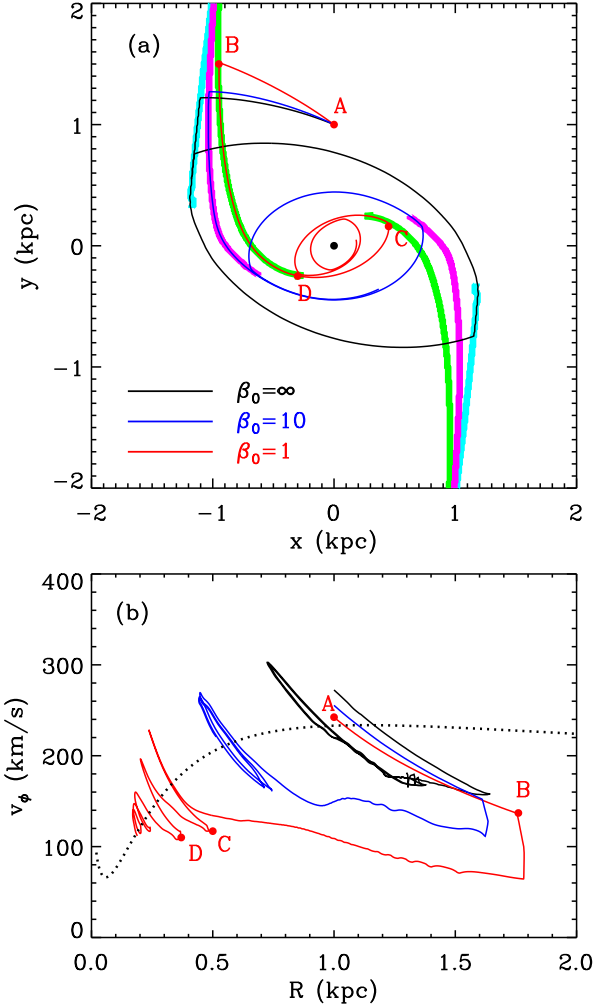} \caption{(a) Instantaneous
streamlines of gas that starts from Point A $(x,y)=(0, 1.5\kpc)$ in
Models bh7MHD01 (red), bh7MHD10 (blue), and bh7HD (black) at $t=150$
Myr. The thick lines nearly parallel to the $y$-axis represent the
dust lane shocks in these models.  In Model bh7MHD01, the gas is
shocked at Point B, subsequently moves in along the dust lane, hits
the opposite-side dust lane at Point C, and returns to the first one
to hit it at Point D.  (b) Variations of the rotational velocity of
the gas along the trajectories shown in (a).  The initial,
equilibrium circular velocity is plotted as a dotted line.
\label{fig:stm}}
\end{figure}

The gas moving on $\xone$-orbits just inside the outermost
$\xone$-orbit collides with that on the 4/1-resonant family,
subsequently switching to lower $\xone$-orbits (e.g.,
\citealt{con89,eng97}; Paper I). Gas on these lower $\xone$-orbits
is continually perturbed by strong total (thermal plus magnetic)
pressure near the bar ends, and after many revolutions it eventually
transits to a trajectory that crosses the dust lane shocks. Figure
\ref{fig:stm}a plots instantaneous streamlines of such gas that
passes through Point A at $(x,y)=(0, 1.5\kpc)$ before hitting a dust
lane in Models bh7MHD01, bh7MHD10, and bh7HD at $t=150$ Myr. The
locations of the dust lanes are indicated as thick lines. The solid
lines in Figure \ref{fig:stm}b plot the change of the rotational
velocity $\vphi$ along the trajectories shown in Figure
\ref{fig:stm}a, while the dotted line draws the rotation curve in
initial equilibrium. The gas hits the shock at Point B and loses a
significant amount of angular momentum there.  Note that in
magnetized models $\vphi$ experiences a dramatic drop at the shock
due largely to magnetic tension forces from the bent field lines.
Coriolis force makes the shocked gas rotate faster gradually as it
moves radially in along the dust lane. In MHD models, magnetic
stress in the dust lanes continuously removes angular momentum from
the inflowing gas, causing $\vphi$ to increase more slowly and thus
making it move further in than in the unmagnetized model.

Figure \ref{fig:stm} shows that in models with $\beta_0\geq10$, the
instantaneous streamlines make an almost closed loop after hitting
the dust lane at the opposite side, forming a nuclear ring in a
position where the centrifugal force balances the gravity (i.e.,
where $\vphi$ achieves its initial value). Due to significant
angular momentum loss at the shocks,  on the other hand, the
streamline in Model bh7MHD01 with $\beta_0=1$ is closed only after
successively hitting the dust lane at the opposite side (Point C)
and then the first one again (Point D), allowing the gas to move in
very close to the galaxy center. With a large supply of gas near the
central parts, magnetized models would have mass inflow rates
greatly enhanced compared to unmagnetized models, as will be shown
in Section \ref{sec:mdot}.

%fig6
\begin{figure}
\epsscale{1.1} \plotone{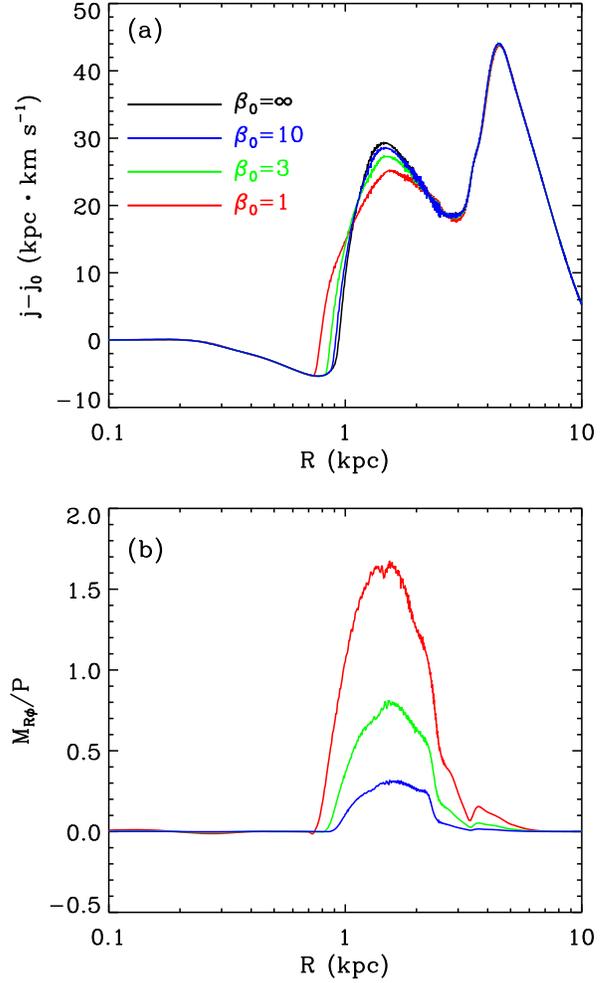} \caption{Radial distributions of
(a) the azimuthally-averaged specific angular momentum $j$ relative
to the initial value $j_0$ and (b) the ratio of the
azimuthally-averaged Maxwell stress $\Max$ to the
azimuthally-averaged thermal pressure $P$ for all bh7 models at
$t=100$ Myr. \label{fig:stress}}
\end{figure}

Is the magnetic stress really responsible for further infall of the
shocked gas toward the galaxy center? To check this, Figure
\ref{fig:stress} plots the radial distribution of the
angle-averaged, specific angular momentum $j =
(2\pi)^{-1}\int_0^{2\pi} R \vphi d\phi$ relative to the initial
value $j_0$ as well as the the ratio of the angle-averaged Maxwell
stress $\Max\equiv (8\pi^2)^{-1}\int_0^{2\pi} B_RB_\phi d\phi$ to
the mean thermal pressure $P=(2\pi)^{-1}\int_0^{2\pi} \cs^2 \Sigma
d\phi$ at $t=100$ Myr for bh7 models. Note that $j > j_0$ at
$R>1\kpc$ for all models.  This is primarily because gas with high
angular momentum at large $R$ has moved inward where $j_0$ is
lower.\footnote{In our models, a portion of the bulge potential is
slowly replaced by the bar potential, with the combined central
density held fixed. A small mismatch between the shapes of the bar
and bulge potentials causes the gas in the central regions to
temporarily expand slightly, leading to $j<j_0$ at $R<1\kpc$.} If
the magnetic stress is important in transporting angular momentum
outward, $j$ would change according to $dj/dt\sim -\cs^2 \Max/P$,
(e.g., \citealt{bal98}). The peak value of the magnetic stress in
Model bh7MHD01 is $\Max/P\sim 1.7$ at $R\sim1.5\kpc$.  Assuming a
constant rate of the angular momentum loss, the net change in $j$
expected over an interval of $\Delta t=100$ Myr is $\Delta j =
4.4\kpc\cdot\kms$, which is very close to the difference in $j$
between Models bh7MHD01 and bh7HD at $R\sim1.5\kpc$. This
demonstrates that the magnetic stress accounts for further loss of
angular momentum and hence a smaller ring size in magnetized models.

%fig7
\begin{figure*}
 \epsscale{1.0} \hspace{-1.5cm}\plotone{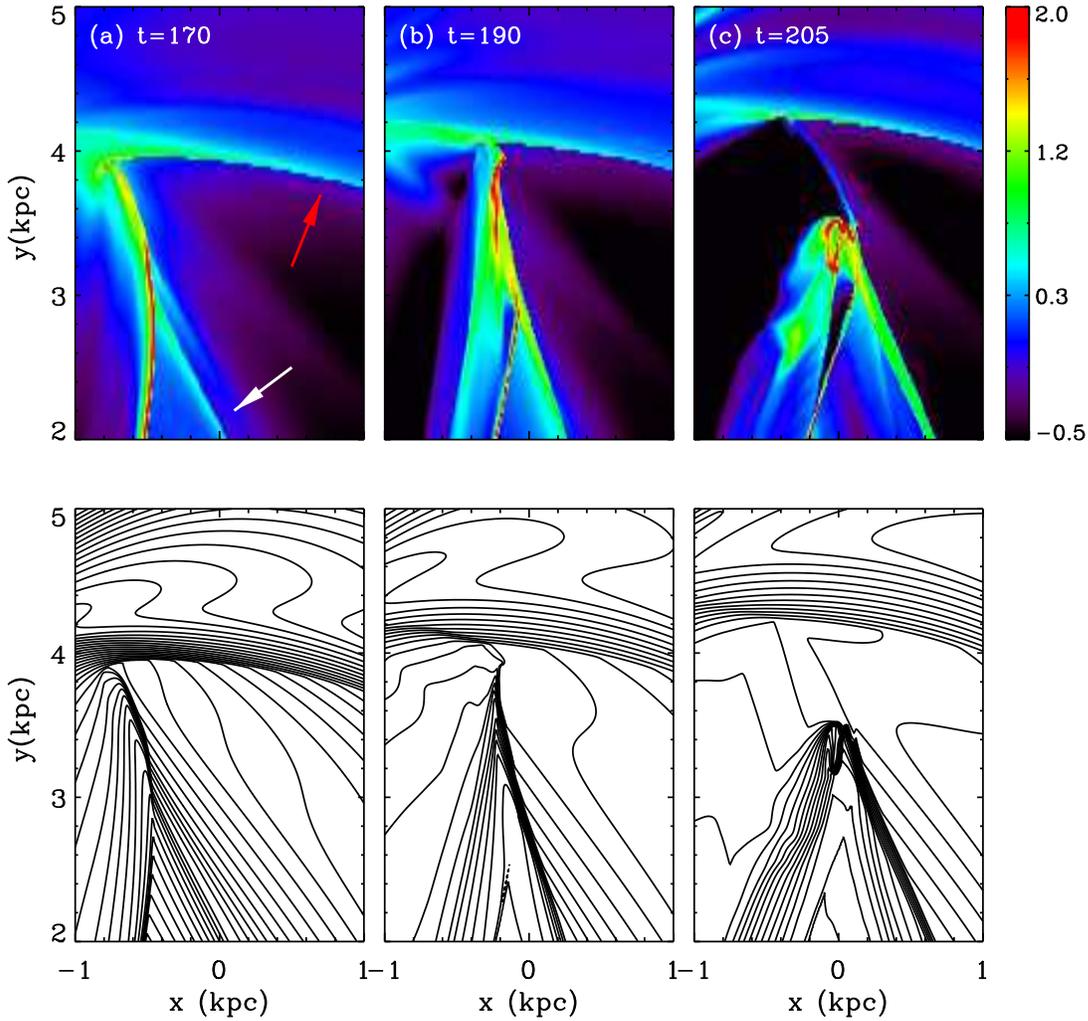}
\caption{Snapshots of logarithm of surface density and the magnetic
field configurations near the upper bar-end regions at $|x|\leq
1\kpc$ and $2\kpc\leq y\leq5\kpc$ of Model bh7MHD01 at (a) $t=170$,
(b) $t=190$, and (c) $t=205$ Myr, illustrating the formation of a
$\Lambda$-shaped magnetic wedge and its detachment from the
outermost $\xone$-orbit. In (a), the red and white arrows mark the
4/1-spiral shocks and smudge, respectively. \label{fig:bend}}
\end{figure*}

The density and magnetic field structures in the bar-end regions are
quite complicated.  Figure \ref{fig:bend} shows a few snapshots of
density and magnetic fields in the upper bar-end regions with
$|x|\leq 1\kpc$ and $2\kpc\leq y\leq5\kpc$ of the standard model.
The 4/1-spiral shock that forms via the collisions of gas on
$\xone$-orbits and on the $4/1$-resonant family is indicated as the
red arrow. Also indicated as the white arrow is a dense ridge of
gas, called ``smudge'', created by the convergence of streamlines at
the backside of the dust lanes (\citealt{pat00}; Paper I).
Compressed magnetic fields in the smudge bend abruptly at the
dust-lane shock,  developing a ``$\Lambda$''-shaped magnetic wedge
that exerts exceedingly large tension force ($t=170$ Myr). At the
same time, the dust lane moves toward the bar major axis and becomes
shorter in extent.  A blob of high-density gas at
$(x,y)\sim(0,3-4\kpc$) near the bar end that was previously
supported by the dust-lane shock becomes loose at $t=190$ Myr, and
starts to move radially in due to the strong tension force in the
magnetic wedge. This produces a strong non-steady flow that rotates
about the galaxy center on $\xone$-orbits just outside the dust
lanes. This non-steady flow, which is stronger in models with
smaller $\beta_0$, dies out slowly as it loses mass near the bar
ends when hitting the dust-lane or 4/1-spiral shocks, with the lost
mass funneled to a nuclear ring along the dust lanes.  At about
$t=500$ Myr, the overall flow pattern reaches a quasi-steady state
where fluid quantities do not change much over time.

\subsection{Dust Lanes}

\begin{figure}
\epsscale{1.1} \plotone{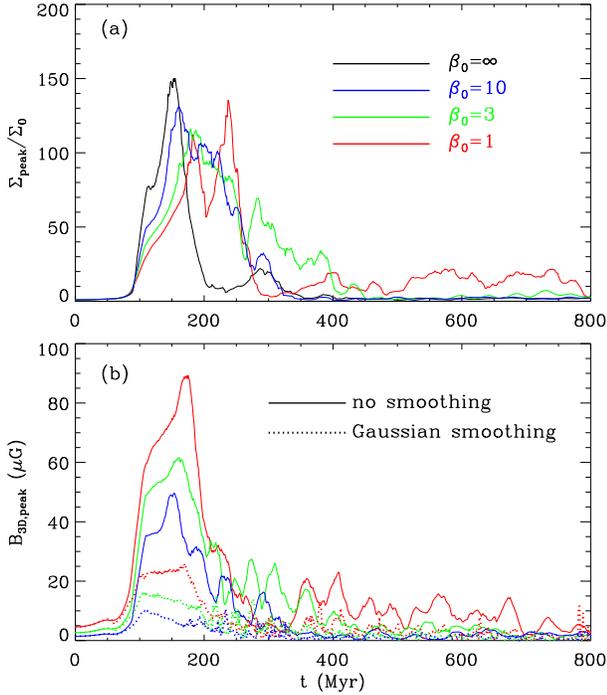} \caption{Temporal evolution of
(a) the peak surface density and (b) peak field strength in the dust
lanes at $y=1.5\kpc$ for bh7 models.  Dotted lines in (b) denote the
field strength obtained by virtual observations with a Gaussian beam
width of $0.1\kpc$. \label{fig:dust}}
\end{figure}

The primary response of gas to the imposed bar potential is the
formation of dust-lane shocks at the leading side of the bar.  Of
course, the degree of density and $B$-field compression depends on
the initial field strength. Figure \ref{fig:dust} plots the temporal
evolution of the peak density and magnetic field of the dust lanes
measured at $y=1.5\kpc$ for bh7 models.  Since the dust lanes are
away from the center, they are unaffected by the BH mass.

%table2
\begin{deluxetable*}{lcccccc}
\tabletypesize{\footnotesize} \tablewidth{0pt}
\tablecaption{Properties of Nuclear Rings\label{tbl:ring}}
\tablehead{ \colhead{Model} & \colhead{$R_{\rm in}$ (kpc)}  &
\colhead{$R_{\rm out}$ (kpc)}  & \colhead{$R_{\rm ring}$ (kpc)}  &
\colhead{$\aSig_{\rm max}/\Sigma_0$} & \colhead{$\Sigma_{\rm
ring}/\Sigma_0$} & \colhead{$\aBth_{\rm max}$ ($\mG$)} } \startdata
bh7MHD01 & 0.03 &  0.71 &  0.34  &  29.5 &  18.6 &  14.8  \\
bh7MHD03 & 0.24 &  0.82 &  0.51  &  29.4 &  20.5 &  20.5  \\
bh7MHD10 & 0.33 &  0.79 &  0.56  &  32.9 &  21.3 &  15.1  \\
bh7HD    & 0.57 &  0.99 &  0.76  &  34.2 &  17.5 &   0.0  \\
\hline
bh0MHD01 & 0.08 &  0.87 &  0.39  &  24.7 &  11.2 &  26.1  \\
bh0MHD03 & 0.25 &  0.81 &  0.51  &  33.8 &  20.8 &  25.8  \\
bh0MHD10 & 0.31 &  0.80 &  0.56  &  32.7 &  21.4 &  13.4  \\
bh0HD    & 0.59 &  1.07 &  0.80  &  32.1 &  15.2 &   0.0
\enddata
\tablecomments{ $R_{\rm in}$ and $R_{\rm out}$ are the inner and
outer radii of the ring defined by the positions where
$\aSig=\aSig_{\rm max}/5$, with $\aSig_{\rm max}$ being the maximum
density; $R_{\rm ring}$ is the mass-weighted ring radius;
$\Sigma_{\rm ring}$ is the mean density of the ring; $\aBth_{\rm
max}$ is the peak strength of magnetic fields inside the ring.}
\end{deluxetable*}

The peak density $\pSig$ of the dust lanes starts to rise rapidly
from $t\sim100$ Myr when the bar potential attains a substantial
amplitude. Magnetic fields reduce the shock compression, making
$\pSig$ smaller with decreasing $\beta_0$. As the gas moves in along
the shocks due to angular momentum loss, the bar region inside the
outermost $\xone$-orbit is progressively evacuated, limiting the
growth of $\pSig$. Models with stronger magnetic fields have dust
lanes closer to the bar major axis.  This makes the gas density in
models with smaller $\beta_0$ more centrally concentrated, causing
$\pSig$ to increase for a longer period of time. In the $\beta_0=1$
model, the second increase of $\pSig$ at $t=200-240$ Myr is caused
mainly by the non-steady gas inflows initiated from near the bar
ends explained above.

Due to shock compression, dust lanes are also regions of strong
magnetic fields.  Figure \ref{fig:dust}b plots the temporal
variations of the peak $B$-field strength in the dust lanes at
$y=1.5\kpc$. In Model bh7MHD01, $\Bpeak$ reaches $\sim90\mG$ before
it declines rapidly as the gas and magnetic fields are advected to
the central regions. Since the dust lanes are extremely narrow with
thickness of $\sim0.02\kpc$ in our models, the observed field
strength of dust lanes is likely to depend sensitively on the
telescope beam size. We illustrate this by plotting as dotted lines
the field strength after Gaussian smoothing with FWHM of $0.1\kpc$,
which is about $\sim4-7$\ times weaker than the value without
smoothing.

%fig8

%fig9
\begin{figure*}
\epsscale{1.1} \plotone{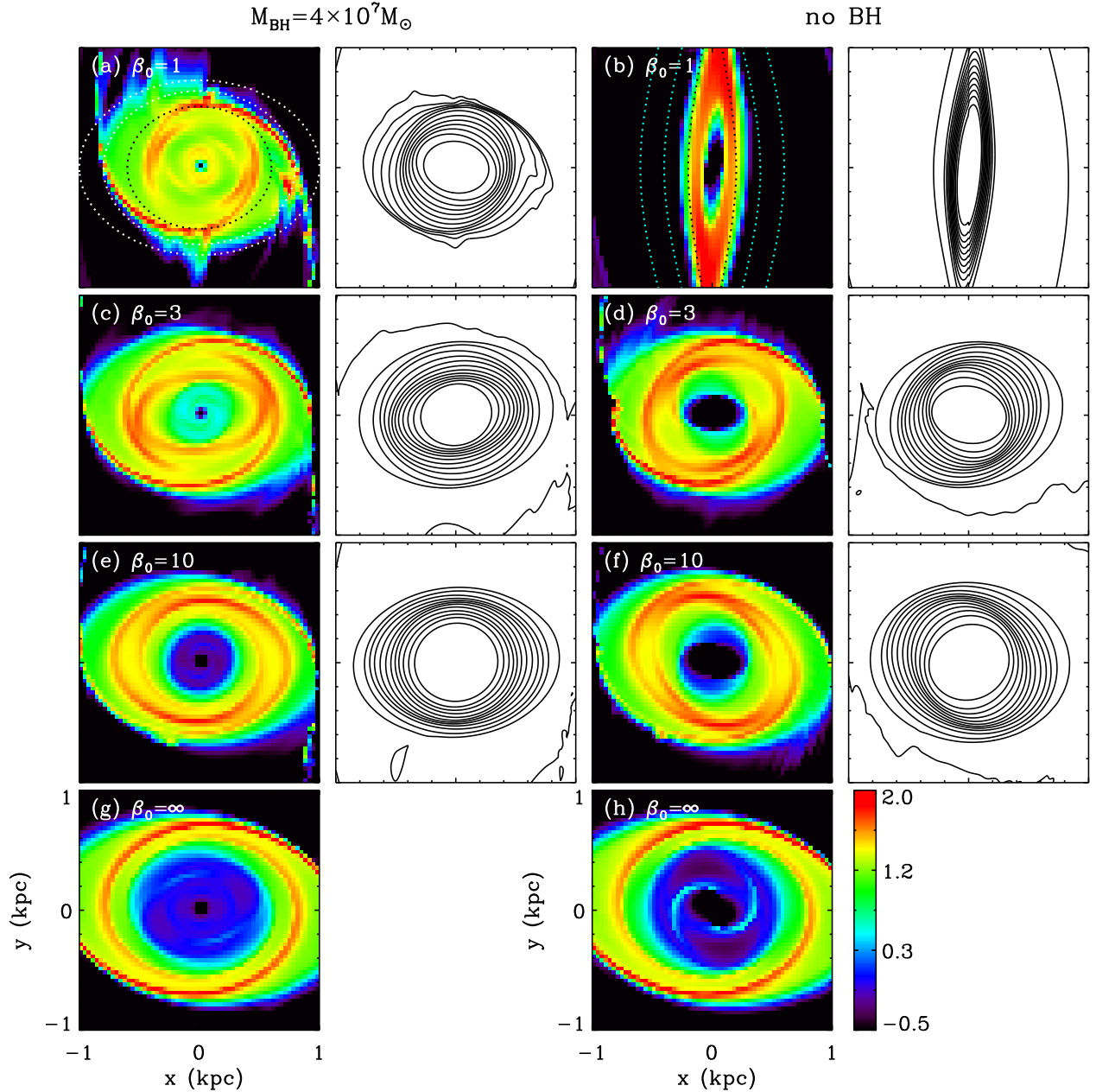} \caption{Logarithm of the density
distribution (color scale) and magnetic field configuration
(contours) in the inner $1\kpc$ regions of all models at $t=800$
Myr. Dotted curves in (a) draw the $\xtwo$-orbits that cut the
$x$-axis at $x_c=0.6, 0.8, 1.0\kpc$, while those in (b) are for the
$\xone$-orbits with $x_c=0.2, 0.4, 0.6\kpc$. \label{fig:allcut}}
\end{figure*}

\subsection{Nuclear Rings}

Figure \ref{fig:allcut} shows the distributions of surface density
and magnetic fields in the central $1\kpc$ regions of all models at
the end of the runs ($t=800$ Myr).  Three  dotted curves  in Figure
\ref{fig:allcut}a represent $\xtwo$-orbits with $x_c = 0.6, 0.8$,
and $1.0\kpc$, while those in Figure \ref{fig:allcut}b are for
$\xone$-orbits with $x_c=0.2, 0.4, 0.6\kpc$. It is apparent that
except for Model bh0MHD01, the shape of a nuclear ring is well
described by $\xtwo$-orbits. Why is the shape of the ring in Model
bh0MHD01 at quasi-steady state so different from the rest?  To
address this question, we first present how the ring forms and
evolves in Model bh7MHD01, and then compare it with the case of
Model bh0MHD01.

%fig10
\begin{figure}
\epsscale{1.2} \plotone{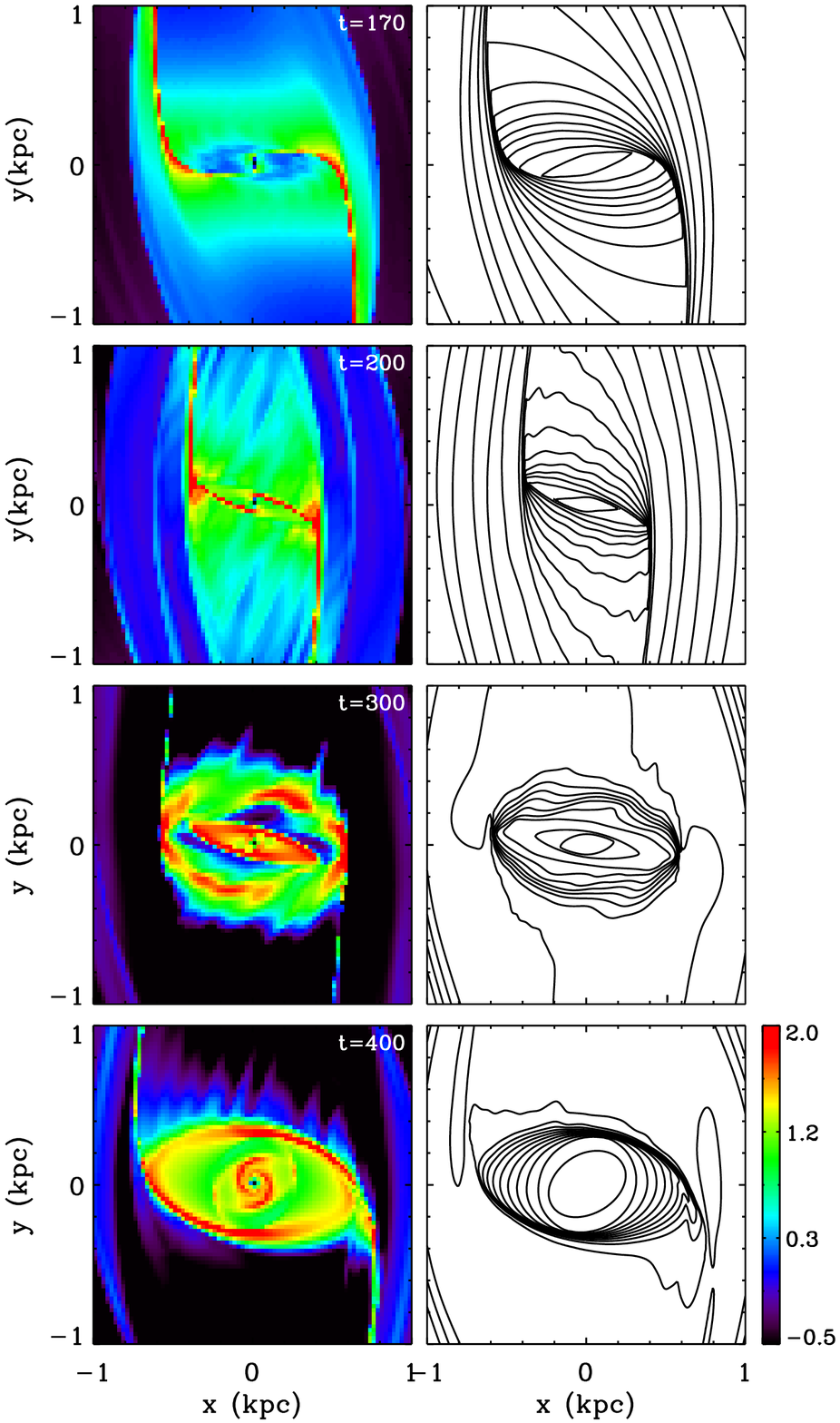} \caption{Snapshots of
logarithm of surface density (left) and field configurations (right)
in the inner $1\kpc$ regions of Model bh7MHD01 at $t=170$, 200,
300, and 400 Myr. \label{fig:m1wh_incut}}
\end{figure}

\subsubsection{Ring Formation}

%fig11
\begin{figure}
\epsscale{1.0} \plotone{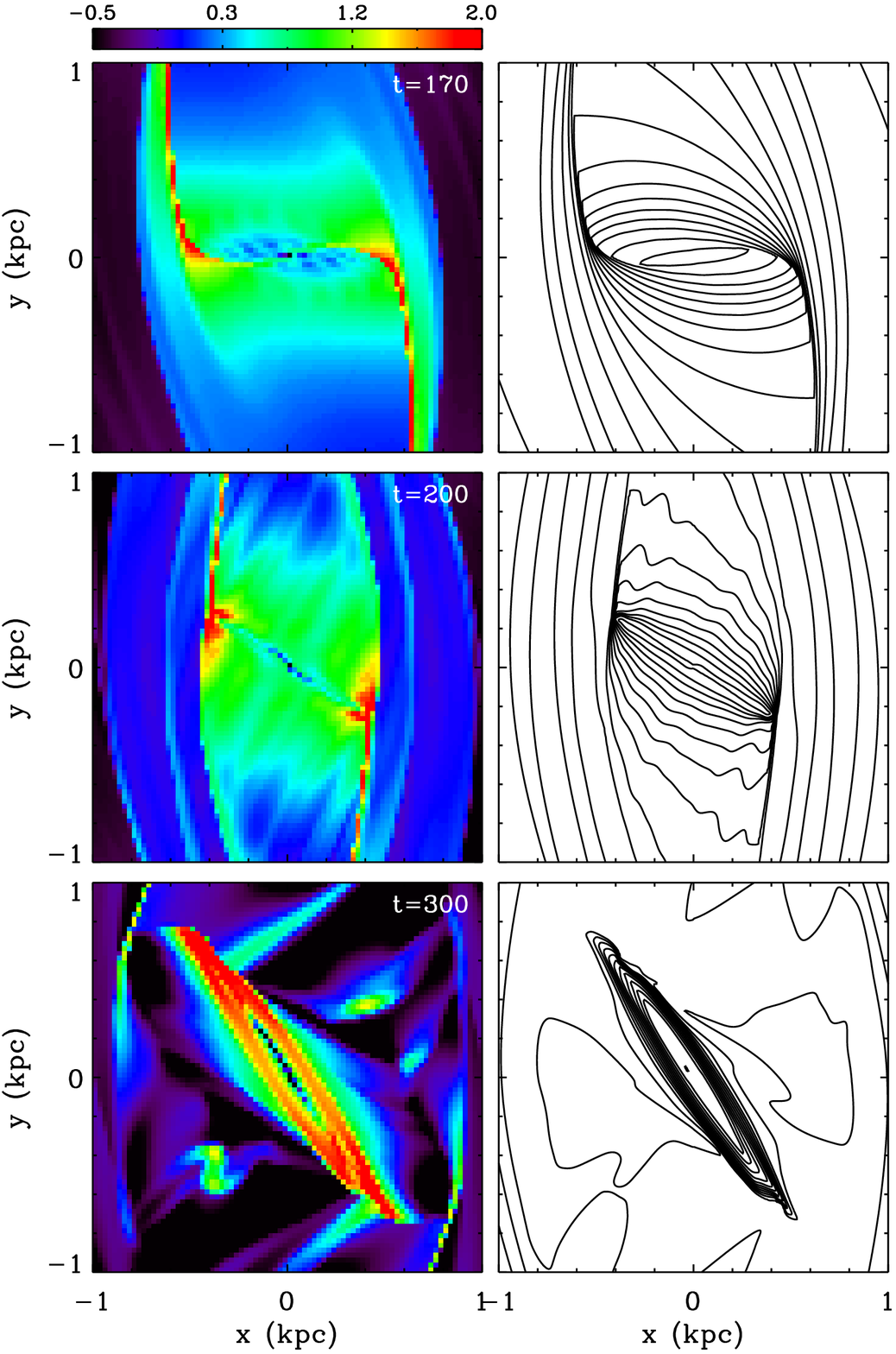} \caption{ Snapshots of
logarithm of surface density (left) and field configurations (right)
in the inner $1\kpc$ regions of Model bh0MHD01 at $t=170$, 200, and
300 Myr. \label{fig:m1h_incut}}
\end{figure}

Figure \ref{fig:m1wh_incut} plots evolutionary changes in gas
morphologies and field configurations in the central parts of Models
bh7MHD01. As the bar potential grows at early time, dust-lane shocks
become stronger and tend to move toward the bar major axis. The
amount of angular momentum loss at the shocks becomes
correspondingly larger with time. At $t=170$ Myr, angular momentum
loss at the shocks due to magnetic tension is so large that the gas
moving in along the dust lanes makes a sharp turn near the $x$-axis
directly toward the galaxy center. This causes the dust lanes to
bend into an ``L'' shape, with the inner (outer) ends aligned
perpendicular (parallel) to the bar major axis. This is unlike in
hydrodynamic models where angular momentum loss at small $R$ is
almost negligible, so that most of the gas moving radially in forms
a ring at $R\sim0.9\kpc$. The inner ends of the L-shaped dust lanes,
which are also shocks, come very close together near the center. In
Model bh7MHD01, they never make contact with each other because of
centrifugal barrier provided by a central BH. Near the center, gas
jumps successively from one inner end to the other at the opposite
side, and thereby keeps losing angular momentum and moving toward
the BH. In a sense, the inner ends of the dust lanes act as ``a bar
within a bar'' proposed by \citet{shl89}, although the inner bar in
our models is a gaseous one.

Both inner and outer dust lanes become stronger with time as the bar
potential continues to grow until $t=186$ Myr. At this time, angular
momentum loss at the outer dust lanes is so large that the gas falls
in toward the BH even before arriving at the $x$-axis, making the
inner ends rotate slightly in the clockwise direction ($t=200$ Myr).
The central gas that hits the inner dust lanes is accreted to the BH
by losing angular momentum further. On the other hand, out-lying gas
whose orbits do not cross the inner dust lanes encircles the center
largely by following $\xtwo$-orbits, gradually transforming to a
nuclear ring. As the amount of gas accreted to the BH increases, the
inner dust lanes become detached from the ring, forming a bar
structure  within the nuclear ring at $t=300$ Myr. This inner,
perpendicular bar feature persists for $\sim100$ Myr. Since the
inner bar is just gaseous structure unsupported by any gravitational
potential, it decays progressively with time, turning into weak
trailing spirals at around $t=400$ Myr. These spirals are
continuously affected by pressure perturbations in the ring,
weakening slowly with time.

Similarly to in Model bh7MHD01, the dust lanes in Model bh0MHD01
with no BH bend into an ``L'' shape at $t=170$ Myr, as shown in
Figure \ref{fig:m1h_incut}. Unlike in the former where a central BH
provides a strong centrifugal force for the inflowing gas, however,
the gas in the latter is almost unresisted and can flow in directly
to the center. This allows the inner ends of the dust lanes to merge
together at the origin ($t=200$ Myr). As a large amount of gas is
accreted to the center during this process, the perpendicular parts
of the dust lanes start to dissolve rapidly with time.

As in Model bh7MHD01, the contact points between the dust lanes and
nuclear ring in Model bh0MHD01 tend to move upstream along the outer
dust lanes. With the inner ends of the dust lanes dissipated away,
the gas near the central part of Model bh0MHD01 follow highly
eccentric orbits that connect the contact points, resulting in an
inclined eccentric ring ($t=180$ Myr). These eccentric orbits are
ballistic, implying that gas dynamical effect is not significant in
the ring.  The inclined ring precesses slowly in the clockwise
direction due to the bar torque that tends to align the semimajor
axis of the ring parallel to the bar major axis. The ring in  Model
bh0MHD01 eventually reaches an equilibrium position at $t=480$ Myr
where the ring gas follows an $\xone$-orbits and feels zero net
torque from the bar.

Less-strongly magnetized models with $\beta_0\geq3$ also form
``L''-shaped dust lanes, with the inner ends connected to each other
at the center when there is no BH. Because angular momentum loss at
the dust-lane shocks is not so significant as in the $\beta_0=1$
model, however, the contact points are always located nearly on the
$x$-axis. The gas orbits connecting the contact points are not so
eccentric as in the $\beta_0=1$ model, either, with their semimajor
axes almost parallel to the $x$-axis.  With negligible net bar
torque, the ring does not precess in these weak $B$-field models. As
the ring continuously interacts with the gas moving in along the
dust lanes, it becomes gradually less eccentric and more distributed
radially,  with its shape remaining similar to an $\xtwo$-orbit.

%fig12
\begin{figure}
\epsscale{1.1} \plotone{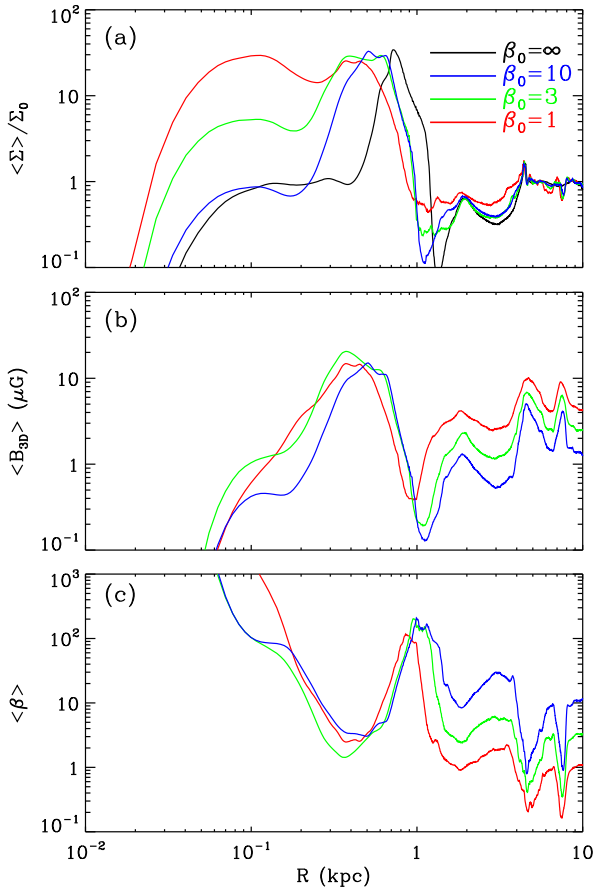} \caption{Radial distribution of
(a) gas surface density, (b) magnetic field strength,
and (c) plasma $\beta$ averaged
both azimuthally and temporally over $t=500-800$ Myr for bh7 models.
\label{fig:ring}}
\end{figure}

\subsubsection{Ring Properties}

Careful inspection of Figure \ref{fig:allcut} reveals that the ring
becomes smaller as $\beta_0$ decreases.  The density and magnetic
fields in the ring are not uniform, often exhibiting weak trailing
spiral features connected to the lower ends of the dust lanes. The
pitch angle of these spirals, notably in Model bh7MHD01, is less
than $10^\circ$.  To quantity the mean properties of the rings,
Figure \ref{fig:ring} plots the radial profiles of surface density
$\aSig$, magnetic fields $\aBth$, and plasma $\langle\beta\rangle$
averaged both azimuthally and temporally over $t=500-800$ yr for all
bh7 models. Table \ref{tbl:ring} gives the inner and outer radii,
$R_{\rm in}$ and $R_{\rm out}$, of the ring, the mass-weighted ring
radius $\Rring=\int_{R_{\rm in}}^{R_{\rm out}} \aSig RdR /
\int_{R_{\rm in}}^{R_{\rm out}} \aSig dR$, the peak density
$\aSig_{\rm max}$, the mean density $\Sigma_{\rm ring}=\int_{R_{\rm
in}}^{R_{\rm out}} \aSig dR/(R_{\rm out}-R_{\rm in})$, and the peak
magnetic field strength $\aBth_{\rm max}$ inside the ring in each
model. Here, $R_{\rm in}$ and $R_{\rm out}$ are defined as the radii
where $\aSig=\aSig_{\rm max}/5$, as in Paper I. The ring becomes
smaller and more centrally distributed due to larger angular
momentum loss in more strongly magnetized models. Small values of
$R_{\rm in}$ in magnetized models imply that a large quantity of gas
is brought in by magnetic torque, readily available for accretion to
the central BH.

In our models, the accretion of magnetic fields to the galaxy center
is realized by removing magnetic fields inside a central hole
together with including the Ohmic dissipation at $R\leq 40\pc$.
Figure \ref{fig:ring}b,c show that the strength of $B$-fields in the
ring ($\aBth\sim15-20\mG$, or equivalently
$\langle\beta\rangle\sim1-3$) in bh7 models does not vary
sensitively to the initial $\beta_0$ since the loss of magnetic flux
via accretion to the galaxy center is larger in models with stronger
initial fields.  While surface density in models with $\beta_0\leq
3$ is more or less uniformly distributed inside $R_{\rm out}$,
magnetic fields peak at $R\sim R_{\rm ring}$ and become weaker at
smaller $R$. The decline of $\aBth$ inward of $R_{\rm ring}$ was
caused by the field accretion as well as the Ohmic dissipation
occurred at $t\sim200-350$ Myr (see Section \ref{sec:mdot}).

\subsection{Nuclear Spirals}\label{sec:nsp}

Figure \ref{fig:allcut} shows that all of the magnetized models do
not possess well-defined nuclear spirals at the end of the runs.
Only Model bh7MHD01 has short spiral-like features that are remnants
of the dissipating inner ends of L-shaped dust lanes.  This is in
contrast to the hydrodynamic results of Paper I that showed that
when the gas is unmagnetized and cold with $\cs=5\kms$, bh0 models
have strong leading spirals, while bh7 models allow weak (but
recognizable) trailing spirals. The absence of nuclear spirals in
magnetized models is primarily due to the fact that rings are
smaller in size and more centrally concentrated than in hydrodynamic
models. The ring material is continuously perturbed by the infalling
gas at the contacted points. The associated thermal pressure and
magnetic forces inhibit growth and maintenance of any coherent
structures in the central parts.

%fig13
\begin{figure*}
\epsscale{1.0} \plotone{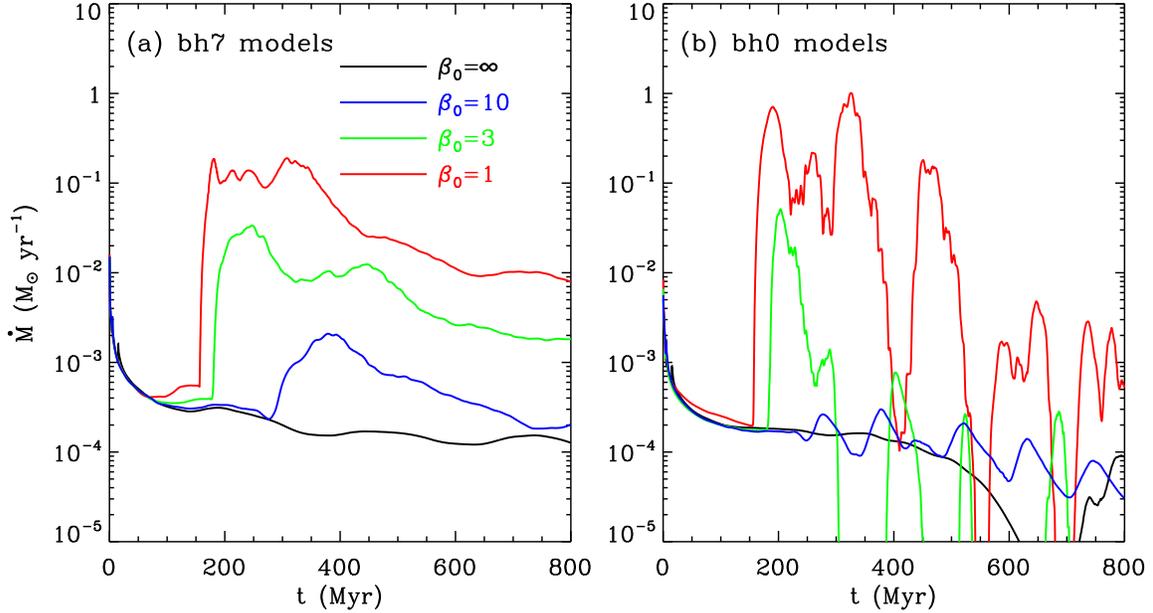} \caption{Temporal variations of
the mass inflow rates $\Mdot$ for (a) models with
$\MBH=4\times10^7\Msun$ and (b) models with no BH.  Magnetized
models with $\beta_0=1$ have $\Mdot$ larger by more than two orders
of magnitude than unmagnetized models. In bh7 models, $\Mdot$ is
sustained, while it becomes relatively intermittent in bh0 models.
\label{fig:mdot}}
\end{figure*}

\subsection{Mass Inflow Rates}\label{sec:mdot}

It has been widely accepted that a galactic bar is an efficient
means to transport the disk gas all the way to the center to fuel a
central BH. Paper I showed that this happens only when the gas has
an effective speed of sound $\cs\geq 15\kms$. In this case, large
thermal pressure at the contact points spreads out the ring material
by perturbing gas orbits away from $\xtwo$-orbits, some of which on
eccentric orbits flow in directly to the center.  On the other hand,
the gas with $\cs\leq 10\kms$ forms a narrow nuclear ring that
prevents further inflows of the gas to the center.

%fig14
\begin{figure}
\epsscale{1.1} \plotone{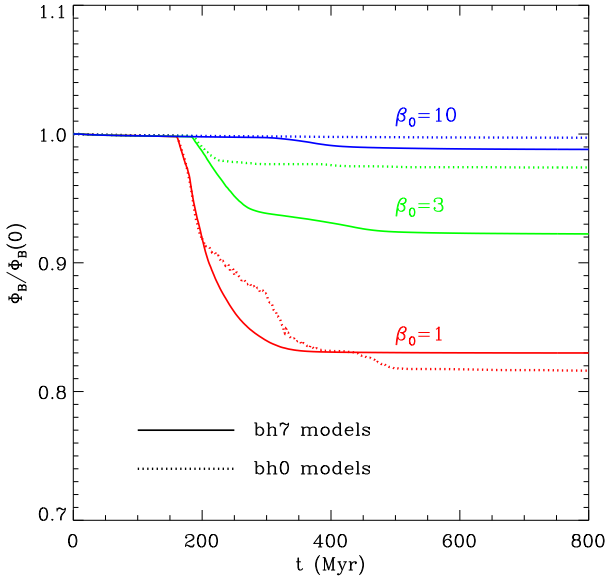} \caption{Temporal variations of
the magnetic flux $\PhiB$ for all models.   The decrease of $\PhiB$
relative to the initial values in bh7 and bh0 models is about 17, 8,
1\% and 18, 3, 0.3\% for $\beta_0=1$, 3, 10 models, respectively.
\label{fig:bflux}}
\end{figure}

%table3
\begin{deluxetable}{lcc}
\tabletypesize{\footnotesize}
\tablewidth{0pt}
\tablecaption{Time-averaged Mass Inflow Rate and Its Dispersion\label{tbl:mdot}}
\tablehead{
\colhead{Model} &
\colhead{$\Mdot  (\Msun\;\yr^{-1})$} &
\colhead{$\Delta \Mdot (\Msun\;\yr^{-1})$}
}
\startdata
bh7MHD01 & $5.1\times 10^{-2}$ & $5.4\times10^{-2}$ \\
bh7MHD03 & $8.5\times 10^{-3}$ & $8.0\times10^{-3}$ \\
bh7MHD10 & $6.5\times 10^{-4}$ & $5.3\times10^{-4}$ \\
bh7HD    & $1.7\times 10^{-4}$ & $4.4\times10^{-5}$ \\
\hline
bh0MHD01 & $8.5\times 10^{-2}$ & $1.8\times10^{-1}$ \\
bh0MHD03 & $1.8\times 10^{-3}$ & $7.1\times10^{-3}$ \\
bh0MHD10 & $1.2\times 10^{-4}$ & $6.3\times10^{-5}$ \\
bh0HD    & $8.6\times 10^{-5}$ & $1.5\times10^{-4}$
\enddata
\tablecomments{
Time average of $\Mdot$ is taken over $t=200-800$ Myr.
}
\end{deluxetable}

We have seen earlier that magnetic stress in the dust lanes removes
a significant amount of angular momentum from the gas in the dust
lanes and thus causes it to move in close to the galaxy center,
potentially increasing the mass inflow rates.  To quantify the
effect of magnetic fields on $\Mdot$, Figure \ref{fig:mdot} plots
temporal changes of $\Mdot$ for all models. The time-averaged value
of $\Mdot$ and its standard deviation $\Delta \Mdot$ over
$t=200-800$ Myr are given in Table \ref{tbl:mdot}. A sudden increase
of $\Mdot$ at $t\sim180$ Myr in $\beta_0=1$ models is due to the
direct inflows of the gas through the inner ends of the L-shaped
dust lanes.  This happens later in models with weaker fields since
it takes the gas longer to reach the galaxy center due to lower
angular momentum loss at the shocks. On average, $\Mdot$ in
$\beta_0=1$ models is larger by more than two orders of magnitude
than in the unmagnetized counterpart. This is of course because
nuclear rings in the former are more centrally concentrated and have
a much larger interior density than the latter.

The presence of a central BH in bh7 models circularizes surrounding
gas orbits and thus makes the initial increase of $\Mdot$ less
dramatic than in the bh0 counterpart where infalling gas can plunge
directly into the central hole.  With more-or-less circular orbits,
the gas in bh7 models flows in continuously to the hole, exhibiting
sustained mass inflows.  On the other hand,  the rapid mass inflows
in bh0 models consume the neighboring gas almost completely,
lowering $\Mdot$ temporarily before fresh gas is supplied from the
outer parts.  When the central regions are filled in, the mass
inflows resume. This makes $\Mdot$ relatively intermittent in bh0
models.\footnote{Assuming a slow and steady inflow due to magnetic
torque, \citet{bec05} estimated $\Mdot\sim1\Mdotunit$, very close to
the peak value in Model bh0MHD01, although mass inflows in our
models occur mostly by gas on non-steady eccentric orbits rather
than steady near-circular ones.}

%fig15
\begin{figure*}
\epsscale{1.1} \plotone{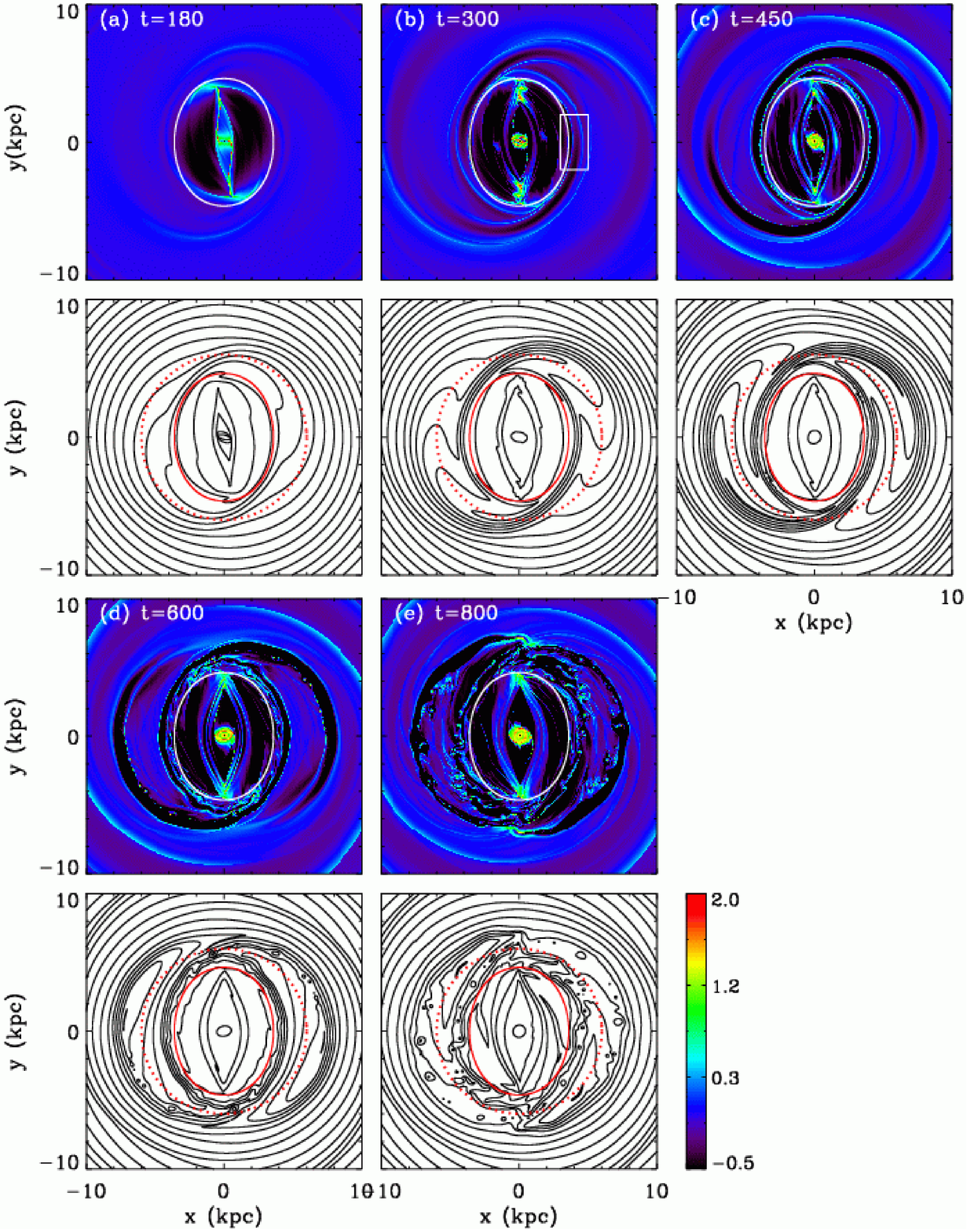} \caption{Snapshots of logarithm
of gas surface density (color scale) and magnetic field
configurations (contours) in the $10 \kpc$ regions of Model bh7MHD01
at $t=180$, 300, 450, 600, and 800 Myr. The solid oval in each panel
draws the outermost $\xone$-orbit, while the dotted circle in the
lower panels with contours marks the CR.  The rectangular section in
(b) is enlarged in Figure \ref{fig:recon}. The MHD dynamo occurring
near the CR produces magnetic arms at $t=450$ Myr.  Magnetic fields
reconnect at the base of the magnetic arms, making density and
$B$-fields chaotic in the outer regions at $t\simgt600$ Myr.
\label{fig:ocut}}
\end{figure*}

To quantify the amount of magnetic fields removed (via accretion or
Ohmic dissipation), we plot in Figure \ref{fig:bflux} temporal
changes of magnetic flux $\PhiB(t)=\int \mathbf{B}\cdot d\mathbf{R}$
relative to the initial value for all magnetized models. A sudden
drop in $\PhiB$ occurs almost at the same epoch as the steep rise of
$\Mdot$, after which $\PhiB$ remains almost constant. This is
because magnetic fields in the central $0.1\kpc$ regions remain very
weak after the rapid accretion phase, so that the magnetic flux
carried in by the gas afterward is not appreciable. The decrease of
$\PhiB$ in bh7 models amounts to $\sim17, 8$, and $1\%$ relative to
the initial value for models with $\beta_0=1, 3$, and 10,
respectively, suggesting that the loss of magnetic flux can be
significant depending on the field strength.

\section{Outer Regions}\label{sec:outer}

As explained in Section \ref{sec:comp}, the outermost $\xone$-orbit
divides a galactic disk essentially into two dynamically
disconnected regions. Gas responses inside it are highly dynamic,
exhibiting dust-lane shocks and significant angular momentum loss
leading to the formation of a nuclear ring. With no crossing of
closed gas orbits, however, the evolution of gas outside the
outermost $\xone$-orbit does not involve shocks and is thus much
less dramatic. In unmagnetized models, the gas at $R>\Rx$
experiences a weak bar torque and moves radially in only slightly,
piling up at the outermost $\xone$-orbit.

%fig16
\begin{figure}
\epsscale{1.2} \plotone{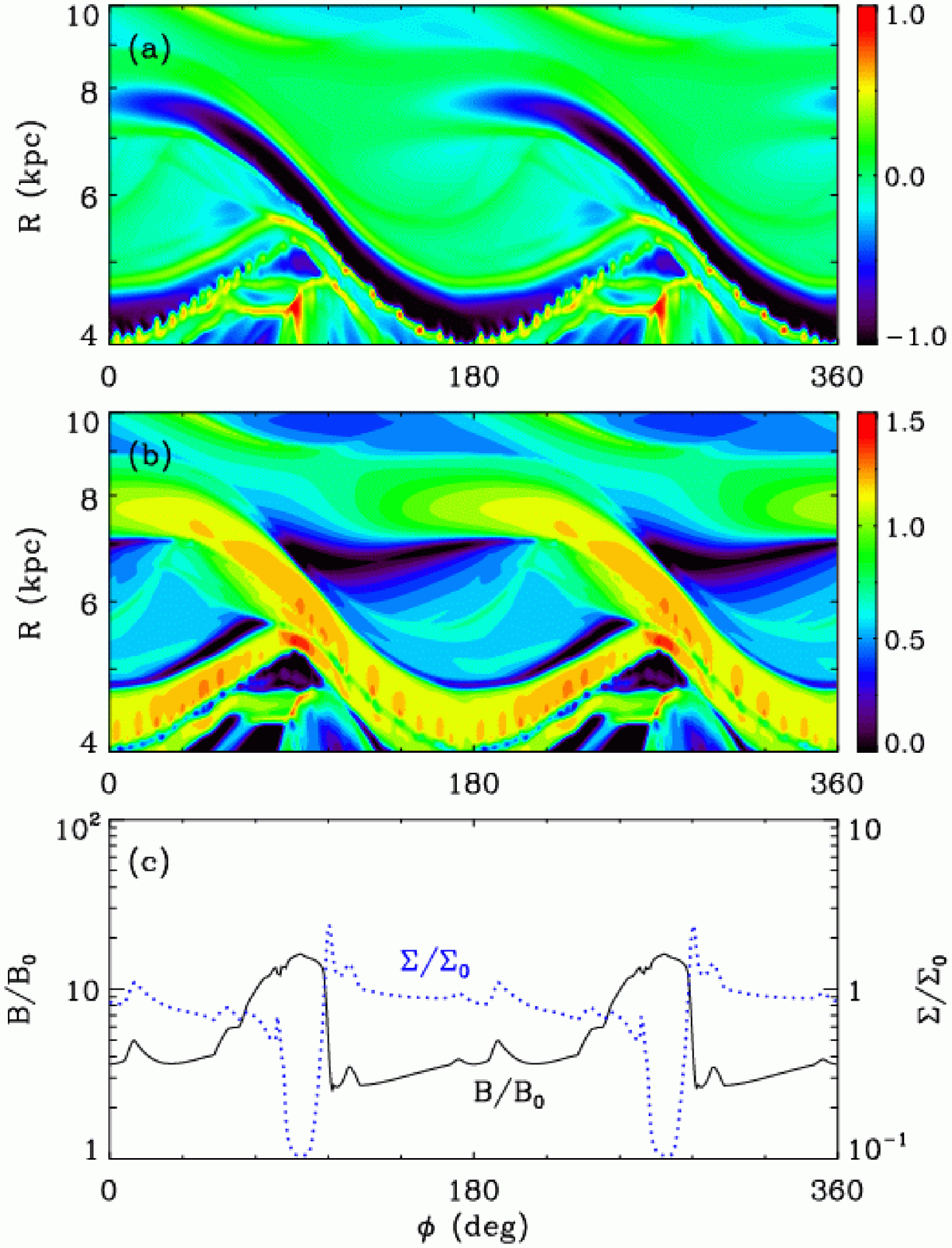} \caption{Logarithm of (a) gas
surface density and (b) magnetic field strength of Model bh7MHD01 at
$t=450$ Myr in the $\ln R$--$\phi$ plane.  The azimuthal
cut-profiles of $\Sigma$ (dotted) and $B$ (solid) at $R=6\kpc$ are
given in (c). \label{fig:logR}}
\end{figure}

The presence of magnetic fields makes the gas responses in the outer
regions completely different from that of unmagnetized cases. Figure
\ref{fig:ocut} shows snapshots of surface density (logarithmic
colorscale) and magnetic fields (contours) in Model bh7MHD01.  The
boxes extend to $10\kpc$ on either side of the center.  The solid
oval in each panel draws the outermost $\xone$-orbit, while dotted
circles in the lower panels with contours indicate the CR. At
$t=180$ Myr, magnetic fields at $R>\Rx$ are relatively unperturbed
compared to those inside.  They closely follow the outermost $
\xone$-orbit near $R=\Rx$ and are slightly compressed along the weak
trailing spirals emerging from the leading side of the bar ends.

Note that the field lines are distorted near the CR and inward of
it. This arises by the combined action of the bar potential and
background shear in such a way that small radial gas motions induced
by the bar potential create the radial component $B_R$ from the
azimuthal component $B_\phi$ of the initial magnetic fields, which
is then stretched by background shear to generate new $B_\phi$ that
is going to turn to $B_R$ again by the radial motions. The whole
process makes a closed cycle, resulting in a secular growth of
magnetic fields in the outer regions. This is an ideal MHD dynamo,
as opposed to the mean-field dynamo, that does not require the
parametric terms in the induction equation. In our models, the MHD
dynamo occurs naturally under a combined action of the bar potential
and galactic shear. Figure \ref{fig:ocut} shows that the MHD dynamo
operates most efficiently near the CR regions along the bar minor
axis.

%fig17
\begin{figure*}
\epsscale{0.9} \plotone{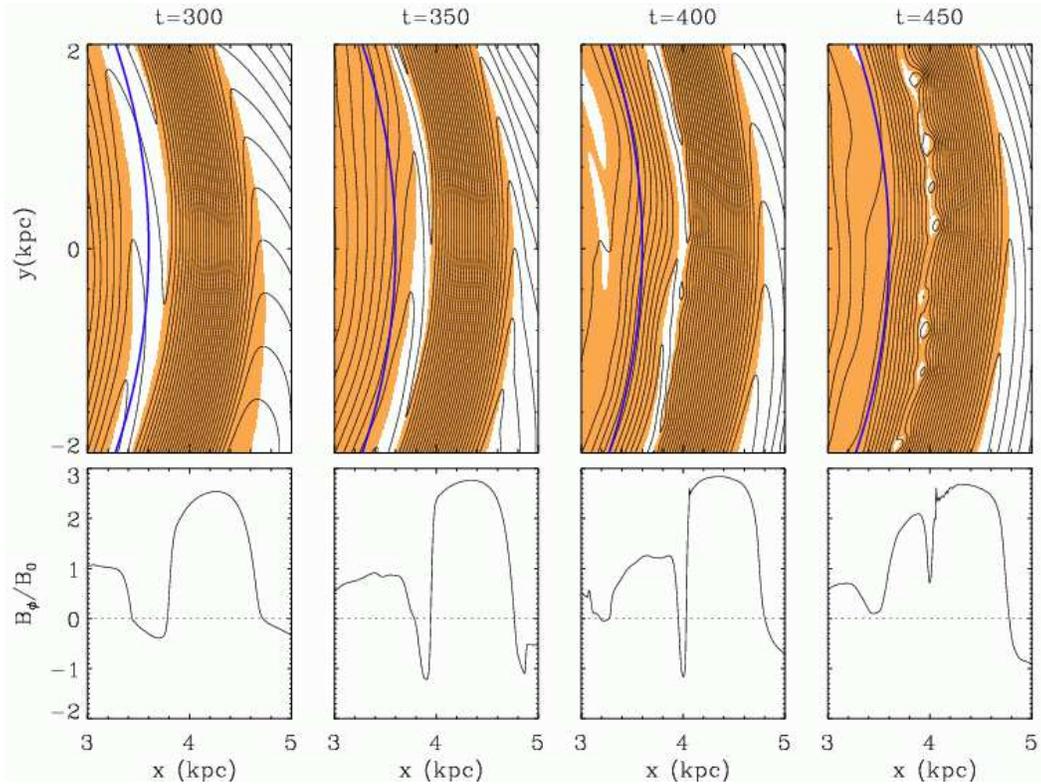} \caption{Upper panels: evolution
of magnetic field configurations in the rectangular section shown in
Figure \ref{fig:ocut}. The shaded regions represent the domains with
positive $B_\phi$, while those with negative $B_\phi$ are unshaded.
Lower panels: profile of $B_\phi$ along the $y=0$ cut.  Magnetic
fields start to reconnect at $t=400$ Myr in a thin current sheet
located at $x\sim4\kpc$ via tearing-mode instability, and develop
numerous magnetic islands along the base of magnetic arms at $t=450$
Myr. \label{fig:recon}}
\end{figure*}

Through magnetic stress, the distorted field lines transport angular
momentum from inside to outside, causing gas and newly generated
field lines at smaller (larger) $R$ to move radially inward
(outward).  Since the field lines do not move across the
outermost-$\xone$ orbit, those moving in are continuously
accumulated there and increase the field strength at the base of the
trailing spiral arms, while those moving outward are swept by galaxy
rotation and added to the outer parts of the arms ($t=300$ Myr).
Although less apparent in Figure \ref{fig:ocut}b, the MHD dynamo is
also effective near the bar ends just inside the outermost
$\xone$-orbit where both $B_R$ and shear are strong: $B_R$ is
stretched by the gas flows just outside the dust lanes, generating
new $B_\phi$ that moves radially outward and is added to the base of
the spiral arms.

Due to the MHD dynamo action occurring near the CR, the spiral arms
become increasingly more magnetized.  The associated strong magnetic
pressure expels the gas away from the arms, eventually forming
``magnetic arms'' that are characterized by stronger magnetic fields
with lower density than the surrounding region at $t=450$ Myr. This
confirms the results of \citet{kul09,kul10} who first found that the
bar potential develops magnetic arms in outer regions. Figure
\ref{fig:logR} plots gas surface density and magnetic field strength
in the $\ln R$--$\phi$ plane as well as the their cut profiles at
$R=6\kpc$ for Model bh7MHD01 at $t=450$ Myr. It is apparent that the
magnetic arms at $R=4.5-7\kpc$ are approximately logarithmic in
shape with a pitch angle of $\sim18^\circ$. The peak $B$-field
strength of the magnetic arms at $R=6\kpc$ is $B_{\rm peak}/B_0=17$,
occurring at the location where gas surface density is minimized at
$\Sigma_{\rm min}/\Sigma_0=0.1$. The gaseous arms produced by the
expelled gas from the magnetic arms have the peak density
$\Sigma_{\rm peak}/\Sigma_0=24$ and lead the magnetic arms by
$\sim15^\circ$ in the azimuthal angle. The magnetic arms keep
growing in strength as the MHD dynamo continues operating.  Since
the added fields are preferentially azimuthal, the outer ends of the
arms curl back in ($t=600$ Myr) and touch the bar end at the
opposite side ($t=800$ Myr).

Figure \ref{fig:ocut}e shows that surface density and magnetic
fields in the outer regions of Model bh7MHD01 become quite chaotic
at the end of the run. This is caused by reconnection of magnetic
fields occurring at the base of the magnetic arms. Figure
\ref{fig:recon} illustrates how magnetic fields reconnect in the
rectangular section shown in Figure \ref{fig:ocut}b from $t=300$ to
450 Myr. The shaded (unshaded) regions in the upper panels represent
domains with positive (negative) $B_\phi$,  with the interfaces
representing the reversal of the field direction. The thick blue
line in the upper panels indicates a part of the outermost
$\xone$-orbit. The lower panels plot the profiles of $B_\phi$ along
the $y=0$ cut. At $t=300$ Myr, the shaded region at the right side
of the outermost $\xone$-orbit corresponds to a segment of the
magnetic arms. The sheared magnetic fields with negative $B_\phi$ to
the left of the magnetic arms are pushed outward as the inner
regions expand due to the pressure gradient ($t=350$ Myr). With the
base of the magnetic arms acting as a rigid wall, this squeezes the
magnetic fields of opposite polarity into a very narrow layer,
developing a thin current sheet at $x\sim4\kpc$ ($t=400$ Myr).
Non-zero numerical resistivity allows the fields to reconnect in the
current sheet through a tearing-mode instability, producing numerous
magnetic islands distributed along the base of the arms ($t=450$
Myr). Magnetic islands that are also sites of density compression
move along the magnetic arms in the course of galaxy rotation, and
interact with the surrounding gas to produce chaotic density and
magnetic structures in the outer regions.

\begin{figure}
\epsscale{1.1} \plotone{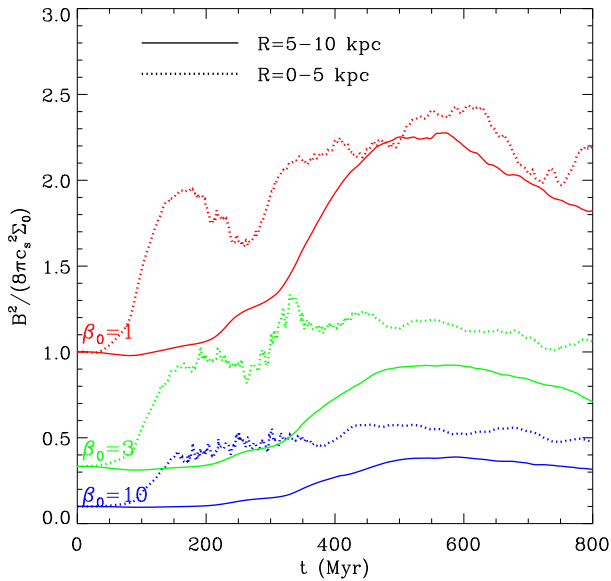} \caption{Temporal changes of the
ratio of magnetic to thermal energy densities in an annulus with
$R=5-10\kpc$ (solid) and in the interior region with $R<5\kpc$
(dotted) in all magnetized models. \label{fig:magE}}
\end{figure}

Figure \ref{fig:magE} plots the temporal evolution of the mean
magnetic energy density $e_m = B^2/(8\pi)$ relative to the initial
thermal energy density for bh7 models.  The solid and dotted lines
are for an annulus with $R=5-10\kpc$ and its interior at
$R=0-5\kpc$, respectively. The initial rise of $e_m$ in the interior
region at $t=50-150$ Myr is due to the formation of dust lanes where
$B$-fields are compressed, while the second rise at $t\sim300-350$
is due to the MHD dynamo occurring at the bar ends. The MHD dynamo
operating near the CR is responsible for the growth of $e_m$ in the
$R=5-10\kpc$ annulus for $t<500$ Myr.  The corresponding increase of
the magnetic energy density is a factor of $\sim2$ for $\beta_0=1$
models and $\sim5$ for $\beta_0=10$ models. The ensuing decrease of
$e_m$ results from the magnetic reconnection that not only reduces
magnetic energy but also makes the outer disk chaotic.

\section{Summary and Discussion}\label{sec:sum}

\subsection{Summary}

We run high-resolution MHD simulations using a modified version of
the Athena code to study the effects of magnetic fields on bar
substructures and the mass inflow rates in barred galaxies. This
work directly extends Paper I in which we explored the case of
unmagnetized disks using the CMHOG code. Most previous studies on
magnetic fields in barred galaxies employed the mean-field dynamo,
ignoring the back reaction of magnetic fields that evolve passively
according to the velocity fields obtained from hydrodynamic runs
(see Section 1 for references). On the other hand,
\citet{kul09,kul10} and \citet{kul11} performed three-dimensional
simulations using ideal MHD models, but they focused mainly on the
formation of magnetic arms in the outer regions and their
morphological changes due to varying rotation frequency and sound
speed.  While our models are two-dimensional, they have about 10
times higher resolution in the in-plane direction than the
three-dimensional models of \citet{kul10}, enabling detailed study
of the bar and nuclear regions.

We consider an infinitesimally-thin, isothermal, rotating,
magnetized gas disk with initially uniform surface density. The
magnetic fields are initially purely azimuthal and uniform with
strength measured by the dimensionless plasma parameter $\beta_0$.
We fix the sound speed to $\cs=5\kms$ and vary $\beta_0$ as well as
the mass $\MBH$ of a BH that controls the rotation curve near the
galaxy center. The main results of the current paper can be
summarized as follows.

1.  Comparisons between the results of hydrodynamic models using the
CMHOG and Athena codes show that except for small differences in the
ring size and location of dust lanes, two results are in good
agreement with each other. The differences are most likely due to
the fact that the Athena runs with a uniform Cartesian grid have
higher spatial resolution at $R>1.1\kpc$, so that the dust-lane
shocks are a bit stronger, resulting in larger angular momentum loss
than in the CMHOG runs that use a non-uniform cylindrical grid. On
the other hand, the CMHOG runs resolve the central regions much
better and thus have stronger nuclear spirals than in the Athena
runs. The close agreement between the results from the difference
codes confirms not only that the bar forces are correctly
implemented in the CMHOG code used in Paper I, but also that the
Cartesian Athena code is reliable in handling dynamics of rotating
flows.

2.  Our adopted model for the external gravitational potential has
an outermost $\xone$-orbit that crosses the $x$- and $y$-axes at
$x_c=3.6$ and $y_c=4.7\kpc$, respectively, and has a Jacobi energy
$E_J=-1.24\times10^5 (\rm km\;s^{-1})^2$, relative to which gas
responses to the bar potential is completely different between
inside and outside. Inside this orbit (i.e., in the bar regions),
there exist families of closed $\xone$- and $\xtwo$-orbit that cross
each other, so that gas on these orbits collides to produce shocks
that eventually develop into dust lanes and a nuclear ring. As the
gas loses angular momentum and moves toward the galaxy center, the
bar regions becomes progressively emptied, decreasing the strength
of the dust lanes. Outside this orbit, on the other hand, there is
no closed orbit and the bar forces are weak, resulting in much
milder gas responses than in the bar regions.

%fig18

3.  Even in the presence of magnetic fields, the bar regions produce
a pair of dust-lane shocks at the leading side of the bar and a
nuclear ring, just like in unmagnetized models.  However, magnetic
fields make several quantitative changes in the properties of bar
substructures and mass inflow rates.  First, magnetic fields
compressed in dust lanes tend to reduce the peak density of the
shocks compared to the unmagnetized counterpart. Second, the
post-shock inflows immediate behind of the shocks along the dust
lanes rotate magnetic fields abruptly at the shock fronts.  The bent
field lines exert magnetic tension forces to the gas moving across
the shocks, removing further angular momentum from it. This causes
the infalling gas to move closer to the galaxy center, forming a
nuclear ring that is smaller in radius and more centrally
distributed than in the hydrodynamic model. Third, small
centrally-concentrated rings in magnetized models destroy coherent
perturbations in the nuclear regions that would otherwise grow into
nuclear spirals in magnetized models.  Fourth, the enhanced density
in the nuclear regions increases the mass inflow rates greatly.  For
instance, magnetized models with $\beta_0=1$ have a time-averaged
mass inflow rate of $\Mdot \sim  (5-8)\times10^{-2}\Mdotunit$, which
is larger by more than two orders of magnitude than that in the
unmagnetized models.

4. In models with $\beta_0\geq3$, a rapid loss of angular momentum
caused by magnetic stress at dust-lane shocks causes them to bend
transiently into an ``L'' shape, with the lower ends pointing
roughly perpendicular to the bar major axis.  The lower ends of the
dust lanes essentially play a role of ``bars within bars" since the
gas moving across them keeps losing angular momentum, moving toward
the galaxy center. Unsupported by any gravitational potential, this
bar-within-a-bar phase does not persist, lasting only for about
$100\Myr$.  As the central gas is lost via accretion to a BH, the
lower ends of the dust lanes dissipate gradually, turning to
trailing nuclear spirals in models with $\MBH=4\times10^7\Msun$
that decay with time.

5. The shape of a nuclear ring at the end of the run can be
completely different depending on the parameters. In our models, all
nuclear rings that form follow an $\xtwo$-orbit closely except for
Model bh0MHD01 (with $\beta_0=1$ and no BH) in which the ring shape is
described rather by an $\xone$-orbit elongated along the bar major
axis. In the latter model, there is no centrifugal barrier
associated with a central BH, so that the lower ends of the L-shaped
dust lanes come very close together to merge at the center, after
which they dissolve rapidly with time. The remaining gas forms a
highly eccentric ring that is inclined with respect to the bar minor
axis.  The inclined ring under the bar torque precesses slowly in the
clockwise direction to align its long axis parallel to the bar major
axis.

6.  In hydrodynamic models, the regions outside the outermost
$\xone$-orbits are almost featureless other than possessing a pair
of weak trailing spirals that emerge from the leading side of the
bar ends. In magnetized models, however, the outer regions are
dynamically active involving an MHD dynamo, magnetic arms, and field
reconnection. The MHD dynamo occurs near the CR and bar-end regions
where both the bar forces and background shear are strong, so that
the bar potential induces the radial velocity perturbations,
producing $B_R$ from $B_\phi$. Background shear then stretches $B_R$
to generate $B_\phi$ that subsequently turns into $B_R$ due to the
radial velocity perturbations, closing the loop of the MHD dynamo
cycle.  The increase of magnetic energy due to the MHD dynamo is
about a factor of $\sim2-5$ with a larger value corresponding to
stronger initial fields. The amplified fields move inward/outward
radially, adding to fields in the trailing spiral arms as well as
their base. As the arms become more magnetized, the arm gas is
expelled by strong magnetic pressure, forming magnetic arms that
have stronger fields but lower density than the surrounding regions.
Magnetic fields with opposite polarity produced by the MHD dynamo
are compressed into a thin layer at the base of the magnetic arms.
With non-zero numerical resistivity, the fields start to reconnect
in the layer via conventional tearing-mode instability, producing
numerous magnetic islands with large density. These magnetic islands
propagate along the magnetic arms to make the outer regions highly
chaotic.

\subsection{Discussion}

Our numerical results show that the density and magnetic fields in
dust lanes remain strong only for 100 Myr around the time when the
bar potential achieves the full strength. The rather rapid decline
of the strength of dust lanes is primarily due to the fact that the
gas located outside the outermost $\xone$-orbit is trapped there and
unable to move further in.  As the gas inside this orbit experiences
shocks and falls in toward the center, the bar regions become
evacuated and the density in the dust lanes drops accordingly.
Observations indicate that most barred galaxies without inner rings
possess a pair of prominent dust lanes (e.g., \citealt{kor04}) that
are nearly straight in strong bars \citep{kna02,com09}. Our results
suggest that barred galaxies with strong dust lanes should have a
stellar bar that is dynamically young, or there should be mechanisms
replenishing the gas in the bar regions.  Candidate mechanisms for
the latter include a spiral-arm potential outside the CR that
perturbs gas orbits to send it in across the outermost $\xone$-orbit
and infalls of halo gas in the form of galactic fountains (e.g.,
\citealt{fra06,fra08}).

Radio polarization observations of barred spiral galaxy NGC 1097
\citep{bec99,bec05} reveal that (1) radio ridges roughly coincide
with the dust lanes in the bar region; (2) the direction of the
magnetic fields changes strongly in the upstream side, making the
polarized emission almost absent there; (3) the equipartition
strength $B_P$ of the regular plus anisotropic random magnetic
fields is $\sim10-20\mu$G in the nuclear ring; (4) $B_P \sim
7-12\mu$G in the dust lanes; (5) the regular fields pass through the
nuclear ring, with an inclination angle of $\sim50^\circ$, which
appears to deviate from the streamlines of the gas flow around the
ring; and (6) the regular fields in the nuclear region have a spiral
shape with large pitch angles.  Points (1)-(3) are entirely
consistent with our numerical results.  Point (4) appears consistent
with our results provided that the bar in NGC 1097 is dynamically
young; otherwise density and magnetic fields of the dust lanes would
not be strong enough to be observed. Again, rejuvenation of the dust
lanes due to gas inflows by spiral arms and/or galactic fountains
would be an alternative possibility. Points (5) and (6) appear
inconsistent with our results that show that magnetic fields follow
the nuclear rings very closely with a pitch angle less than
$10^\circ$ and nearly circular in the central parts (see Fig.\
\ref{fig:allcut}). These discrepancies may owe to the fact that our
models are limited to a two-dimensional razor-thin geometry with
purely azimuthal initial fields, and thus are unable to capture the
potential effects of poloidal fields and other dynamical processes
that may be important in three dimensions. In addition, star
formation and associated small-scale turbulent dynamo occurring in
nuclear rings of real galaxies may considerably affect the field
configurations in the central regions.

The concept of ``bars within bars'' was introduced by \citet{shl89}
to overcome the difficulty, caused by the formation of a narrow
nuclear ring, of a single large-scale bar in bringing the disk gas
all the way to within $\sim 1\pc$ from the galaxy center to feed an
AGN. Near-infrared and/or CO observations of nearby barred galaxies
indeed show that a substantial fraction of disk galaxies exhibit
double bars (e.g., \citealt{sha95,fri96,erw99,gar98,mac00}), with
the extent of the secondary (nuclear) bars typically 5 to 7 times
shorter than that of the primary bar. The origin of secondary bars
is yet unclear. They may arise from gravitational instability of a
disk containing both stars and gas (e.g.,
\citealt{fri93,com94,hel94}), or be due to crowding of stellar
orbits perturbed by gravity of a nuclear ring (e.g.,
\citealt{sha93}), or be produced by an unstable gaseous disk in the
central part (e.g., \citealt{shl89}; see also \citealt{ath00} and
references therein). Using a dynamically-possible doubly-barred
galaxy model, \citet{mac02} showed that a hydrodynamic response of
the gas to a self-consistent, secondary-bar potential is much weaker
than that to the primary bar and is thus unable to enhance the mass
inflow rates much. Our numerical results show that magnetic fields
naturally produce, albeit transiently, a bar-within-bar structure
that greatly enhances the mass inflow rates even without invoking
the secondary stellar bar potential. In our models, the secondary
bar (or inner dust lanes) is almost corotating with the outer
primary bar and lasts only for $\sim 100$ Myr before turning to weak
trailing spirals. Since we include the effect of stars via a fixed
gravitational potential, we are of course unable to consider the
back reaction of stars to the gaseous gravity.

Our models show that MHD dynamo action due to a bar potential
combined with background shear produces magnetic arms in the outer
regions, consistent with the results of \citet{kul09,kul10} and
\citet{kul11} who reported that magnetic arms drift into interarm
regions due to a lower angular velocity than gaseous spirals.  In
most external disk galaxies, magnetic field directions based on
polarized synchrotron radiation follow optical spiral structures
fairly well, with stronger total (regular plus turbulent) fields
inside the arms than outside (e.g., \citealt{bec96,fle11}). However,
there are some exceptional galaxies such as IC 342 (e.g.,
\citealt{kra93}) and NGC 6946 (e.g., \citealt{bec_hoe96}) that are
known to have strongest magnetic fields at magnetic spiral arms that
lie in between optical arms. Proposed mechanisms for magnetic arms
include MHD density waves in two dimensions \citep{fan96,lou98} and
mean-field turbulent dynamo \citep{mos98,shu98,roh99}. The MHD wave
theory invokes a systematic phase-shift between fast and slow MHD
modes responsible for optical and magnetic arms, respectively
\citep{lou98}, although this requires rigid-body rotation over a
wide range of radii.  In addition, MHD waves are subject to a
buoyant instability when the vertical degree of freedom is allowed
\citep{shu05}. On the other hand, the mean-field dynamo theory
argues that optical arms have stronger turbulent motions hence
larger turbulent magnetic diffusivity, resulting in weaker fields
than interarm regions, although it relies on uncertain parameters
including the dynamo number and ignores the dynamical effects of
magnetic fields on the gas.

Can the MHD dynamo occurring in the outer regions of our numerical
models account for observed magnetic arms in IC 342 and NGC 6949? We
think this is unlikely since the connection between bar-induced
magnetic arms and those observed is quite uncertain. First of all,
the disk galaxies with observed well-defined magnetic arms are not
strongly barred (e.g., \citealt{reg95}). In addition, the spiral
arms in NGC 6949 have weak, $m=4$ modes, whereas the bar potential
preferentially amplifies $m=2$ arms. Moreover, it is uncertain if
the MHD dynamo in three dimensions occurs similarly to in
two-dimensional in-plane geometry, studied in this work.
Nevertheless, the MHD dynamo as a mechanism for field amplification
in barred galaxies is quite attractive in that it uses only natural
ingredients (bar potential and galactic shear) without making any
assumption.  It may have something do with relatively weak magnetic
arms observed in outer regions of barred galaxies such as NGC 1365
\citep{bec02}, NGC 1097 and NGC 1365 \citep{bec05}, although it is
questionable if the gaseous arms in these galaxies are driven solely
by a bar potential. While our models considered only a bar potential
as a perturbing agent for simplicity, real galaxies also have spiral
arms in the regions outside a bar. It would be interesting to study
how the MHD dynamo due to a bar potential conspires with a
spiral-arm potential to generate and shape magnetic fields in outer
regions.

\acknowledgments We are grateful to W.-Y.\ Seo for help in
implementing the bar potential in the Athena code.  We also
acknowledge helpful comments from R.\ Beck and a thoughtful report
from the referee.  This work was supported by the National Research
Foundation of Korea (NRF) grant funded by the Korean government
(MEST), No.\ 2010-0000712.

\appendix

\section{Numerical Viscosity and Magnetic Diffusivity of the Athena Code}\label{sec:appen}

In this Appendix we evaluate numerical viscosity and magnetic
diffusivity of the Athena code by using the damping rates of
traveling magnetosonic waves. We begin by writing the MHD equations
for isothermal, viscous, and resistive plasma
\begin{equation}\label{eq:con2}
\frac{\partial\rho}{\partial t}
+\nabla\cdot(\rho \mathbf{v}) = 0,
\end{equation}
\begin{equation}\label{eq:mom2}
\frac{\partial \mathbf{v}}{\partial t}
+ \mathbf{v} \cdot \nabla \mathbf{v} = - \frac{\cs^2}{\rho}\nabla \rho
+ \frac{1}{4\pi\rho} (\nabla \times \mathbf{B}) \times \mathbf{B}
+ \nu \nabla^2 \mathbf{v},
\end{equation}
\begin{equation}\label{eq:ind2}
\frac{\partial \mathbf{B}}{\partial t} = \nabla\times
(\mathbf{v} \times \mathbf{B}) + \eta \nabla^2 \mathbf{B},
\end{equation}
where $\nu$ and $\eta$ are the coefficient of kinematic viscosity
and magnetic diffusivity, respectively.  Other symbols have their
usual meanings.

For the purpose of measuring the numerical values of $\nu$ and
$\eta$ of the Athena code used in this work, we limit ourselves to
magnetosonic waves propagating along the direction perpendicular to
the initial magnetic fields $\mathbf{B}_0 = B_0 \mathbf{\hat y}$
through an initially static, uniform medium with density $\rho_0$.
We consider small-amplitude perturbations $\rho_1(x,t)$,
$\mathbf{v}_1 = v_1(x,t) \mathbf{\hat x}$, and $\mathbf{B}_1 =
B_1(x,t) \mathbf{\hat y}$ to the density, velocity, and magnetic
fields, respectively. Assuming $|\rho_1|/\rho_0 \ll 1$,
$|v_1|/\cs\ll1 $ and $|B_1|/B_0 \ll1$, we linearize equations
(\ref{eq:con2}) -- (\ref{eq:ind2}) to obtain a set of the perturbed
equations
\begin{equation}\label{eq:con3}
\frac{\partial\rho_1}{\partial t}  =
-\rho_0 \frac{\partial v_1}{\partial x},
\end{equation}
\begin{equation}\label{eq:mom3}
\frac{\partial v_1}{\partial t} = - \frac{\cs^2}{\rho_0}
\frac{\partial\rho_1}{\partial x}
- \frac{B_0}{4\pi\rho_0} \frac{\partial B_1}{\partial x}
+ \nu \frac{\partial^2 v_1}{\partial x^2},
\end{equation}
\begin{equation}\label{eq:ind3}
\frac{\partial B_1}{\partial t} = -B_0 \frac{\partial v_1}{\partial x}
+ \eta \frac{\partial^2 B_1}{\partial x^2}.
\end{equation}

We now seek for the plane-wave solutions $\rho_1, v_1, B_1 \propto
\exp(ikx - i\omega t)$ with wavenumber $k$ and frequency $\omega$.
Plugging these into equations (\ref{eq:con3}) -- (\ref{eq:ind3}) and
eliminating $\rho_1$ and $B_1$ in favor of $v_1$, we obtain the
dispersion relation
\begin{equation}\label{eq:disp}
\omega^3 + i (\nu+\eta)k^2 \omega^2 -
[(\cs^2 + \vA^2)k^2 + \nu\eta k^4]\omega - i \eta \cs^2k^4 = 0,
\end{equation}
where $\vA=B_0/(4\pi \rho_0)^{1/2}$ is the Alfv\'en speed.
It is trivial to show that when $\nu=\eta=0$,
equation (\ref{eq:disp}) is reduced to the relation
\begin{equation}\label{eq:disp2}
\omega^2 = \omega_0^2 \equiv (\cs^2 + \vA^2) k^2,
\end{equation}
for magnetosonic waves.  The corresponding (real) eigensolutions are
\begin{equation}\label{eq:eigen}
\left(
\begin{array}{c}
\rho_1 \\
v_1 \\
B_1
\end{array}\right)
= \mathcal{A}_0
\left(
\begin{array}{c}
\rho_0     \\
\omega_0/k \\
B_0
\end{array}\right)
\sin(kx-\omega_0t),
\end{equation}
where $\mathcal{A}_0$ is the amplitude of the perturbations in
the absence of viscosity and magnetic diffusivity.

When $\nu$ and $\eta$ are non-zero but small (i.e.,
$\nu,\eta\ll \omega_0/k^2$), equation (\ref{eq:disp}) yields
\begin{equation}\label{eq:omega}
\omega = \omega_0 - i[\nu + \vA^2/(\cs^2+\vA^2) \eta] k^2/2,
\end{equation}
implying that viscosity and magnetic diffusivity cause
the perturbation amplitude in equation (\ref{eq:eigen})
to decay exponentially with time as
\begin{equation}\label{eq:amp}
\mathcal{A}(t) = \mathcal{A}_0 \exp\left[-
\frac{t}{2}\left(\nu
+ \frac{\vA^2}{\cs^2+\vA^2} \eta\right) k^2
\right].
\end{equation}

\begin{figure}
\epsscale{0.8} \plotone{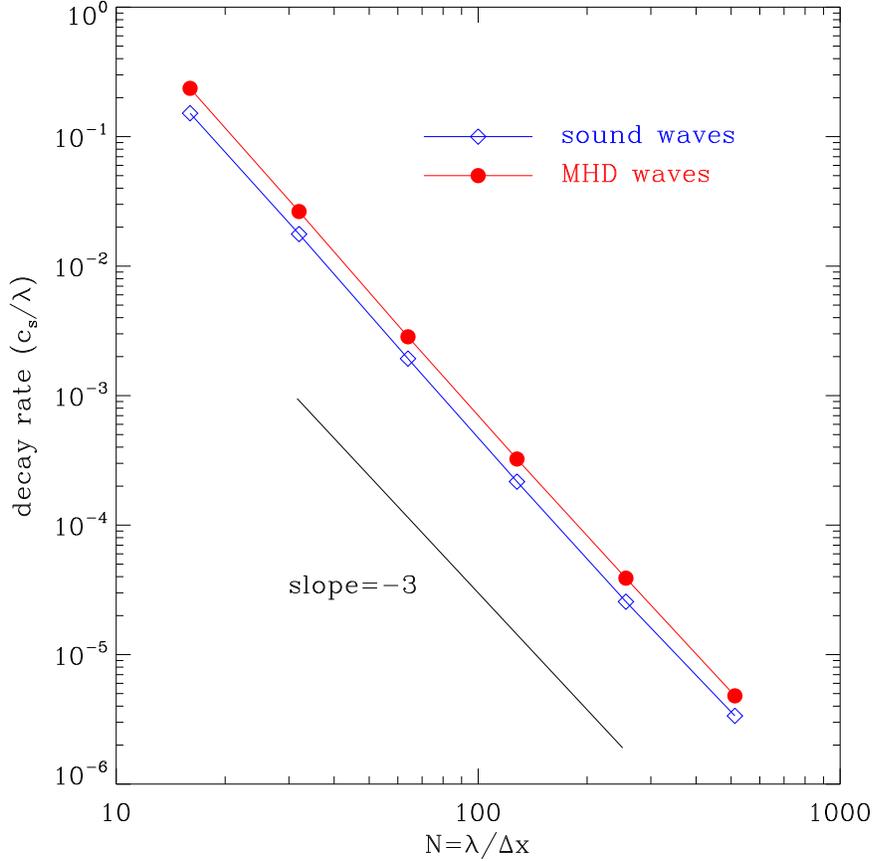} \caption{ Decay rates of acoustic
and MHD waves in the Athena test runs as functions of
$N=\lambda/\Delta x$, the number of grid points per wavelength.
\label{fig:decay}}
\end{figure}

To evaluate $\nu$ and $\eta$ separately, we first consider pure
sound waves with wavelength $\lambda=2\pi/k$ in an unmagnetized
medium, and initialize in the Athena code the perturbations
according to equation (\ref{eq:eigen}) at $t=0$.  We fix the initial
amplitude to $\mathcal{A}_0=10^{-3}$, and monitor the temporal decay
of the wave amplitude. The resulting decay rates for various runs
with differing $N=\lambda/\Delta x$, the number of grid points per
wavelength, are plotted in Figure \ref{fig:decay} as open diamonds.
The best fit to the numerical viscosity is found to be
\begin{equation}\label{eq:vis}
\nu_n = 2.1\times 10^{-7}
\left(\frac{\cs}{5\kms}\right)
\left(\frac{\Delta x}{10\pc}\right)^3
\left(\frac{\lambda}{1\kpc}\right)^{-2}\;\kpc^2\Myr^{-1}.
\end{equation}
Note that $\nu_n$ is proportional to $\Delta x^3/\lambda^2$,
completely analogous to the behavior of the numerical conductivity
of the Athena code reported by \citet{kim08}.

Next, we set up magnetosonic waves using equation (\ref{eq:eigen})
at $t=0$ together with $\mathcal{A}_0=10^{-3}$ and $\cs=\vA$.
Figure \ref{fig:decay} plots as filled circles the numerical damping
rates of the magnetosonic waves due to both viscosity and magnetic
diffusivity. After correcting for the viscous contribution using
equation (\ref{eq:vis}), the residual damping rates are fitted by
the numerical diffusivity as
\begin{equation}\label{eq:diff}
\eta_n = 3.5\times 10^{-7}
\left(\frac{\vA}{5\kms}\right)
\left(\frac{\Delta x}{10\pc}\right)^3
\left(\frac{\lambda}{1\kpc}\right)^{-2}\;\kpc^2\Myr^{-1},
\end{equation}
for $\vA\sim\cs$. Note that $\eta_n$ and $\nu_n$ are of the same
order. While $\eta_n$ is quite small, the related magnetic diffusion
time $\tau_{\rm mag} \equiv \Delta x^2/\eta_n$ can be comparable to
the dynamical time (e.g., bar orbital period) if the field strength
changes substantially across the grid spacing $\Delta x$. For
instance, $\tau_{\rm mag} \sim 100\Myr$ for $\Delta x=7.3\pc$ and
$\lambda\sim0.5\kpc$ typical for tearing modes of magnetic
reconnection occurring in our simulations.

\hspace{0.5cm}


\begin{thebibliography}{}
\bibitem[Ann \& Thakur(2005)]{ann05}
  Ann, H. B., \& Thakur, P. 2005, \apj, 620, 197
\bibitem[Athanassoula(1992a)]{ath92a}
  Athanassoula,~E. 1992a, \mnras, 259, 328
\bibitem[Athanassoula(1992b)]{ath92b}
  Athanassoula,~E. 1992b, \mnras, 259, 345
\bibitem[Athanassoula(2000)]{ath00}
  Athanassoula,~E. 2000, in Alloin D.\ et.\ al,
  ASP Conf.\ Series, V. 221, ed.\ D.\ Alloin, K.\ Olsen, \& G.\ Galaz
  (San Francisco: ASP), 243
\bibitem[Balbus \& Hawley(1998)]{bal98}
  Balbus, S.\ A., \& Hawley, J.\ F.\ 1998, Rev.\ Mod.\ Phy., 70, 1
\bibitem[Beck(2009)]{bec09}
   Beck, R.\ 2009, Astrophys.\ Space Sci.\ Trans., 5, 43
\bibitem[Beck \& Hoernes(1996)]{bec_hoe96}
   Beck, R., \& Hoernes, P.\ 1996, Nature, 379, 47
\bibitem[Beck et al.(1996)]{bec96}
   Beck, R., Brandenburg, A., Moss, D., Shukurov, A., \& Sokoloff, D.\
   1996, \araa, 34, 155
\bibitem[Beck et al.(1999)]{bec99}
  Beck, R., Ehle, M., Shoutenkov, V., Shukurov, A., Sokoloff, D.\
  1999, Nature, 397, 324
\bibitem[Beck et al.(2002)]{bec02}
  Beck, R., Shoutenkov, V., Ehle, M., et al.\ 2002, \aap, 391, 83
\bibitem[Beck et al.(2005)]{bec05}
  Beck, R., Fletcher, A., Shukurov, A., Snodin, A., Sokoloff, D.~D.,
  Ehle, M., Moss, D., \& Shoutenkov, V.\ 2005, \aap, 444, 739
\bibitem[Buta(1986)]{but86}
  Buta, R.\ 1986, \apjs, 61, 609
\bibitem[Buta \& Combes(1996)]{but96}
  Buta, R., \& Combes, F.\ 1996, Fund. Cosmic Phys., 17, 95
\bibitem[Camenzind \& Lesch(1994)]{cam94}
  Camenzind, M., \& Lesch, H.\ 1994, \aap, 284, 411
\bibitem[Combes(1994)]{com94}
  Combes, F.\ 1994, in Mass-Transfer Induced Activity in Galaxies, ed.\
  I.\ Shlosman (Cambridge: Cambridge Univ.\ Press), 170
\bibitem[Comer\`on et al.(2009)]{com09}
  Comer\`on, S., Mart\`inez-Valpuesta, I., Knapen, J.\ H., \& Beckman, J.\ E.\
  2009, \apj, 706, L256
\bibitem[Contopoulos \& Grosb{\o}l(1980)]{con89}
  Contopoulos, G., \& Grosb{\o}l, P.\ 1989, \aapr, 1, 261
\bibitem[Englmaier \& Gerhard(1997)]{eng97}
  Englmaier, P., \& Gerhard, O.\ 1997, \mnras, 287, 57
\bibitem[Erwin \& Sparke(1999)]{erw99}
  Erwin, P., \& Sparke, L.\ S.\ 1999, in ASP Conf.\ Ser.\ 182,
  Galaxy Dynamics, ed.\ D.\ R.\ Merritt \& J.\ A.\ Sellwood
  (San Francisco: ASP)
\bibitem[Fan \& Lou(1996)]{fan96}
  Fan, Z., \& Lou, Y.\ Q.\ 1996, Nature, 383, 800
\bibitem[Ferrers(1887)]{fer87}
  Ferrers, N.\ M.\ 1887, Q.J.Pure Appl.\ Math., 14, 1
\bibitem[Fletcher et al.(2011)]{fle11}
  Fletcher, A., Beck, R., Shukurov, A., Berkhuijsen, E.\ M.,
  \& Horellou, C.\ 2011, \mnras, 412, 2396
\bibitem[Fraternali \& Binney(2006)]{fra06}
  Fraternali F., Binney J.\ 2006, \mnras, 366, 449
\bibitem[Fraternali \& Binney(2008)]{fra08}
  Fraternali F., Binney J., 2008, \mnras, 386, 935
\bibitem[Friedli et al.(1996)]{fri96}
  Friedli, D., Wozniak, H., Rieke, M., Martinet, L., \& Bratschi, P.\ 1996,
  \aaps, 118, 461
\bibitem[Friedli \& Benz(1993)]{fri93}
  Friedli, D., \& Benz, W.\ 1993, \aap, 268, 65
\bibitem[Garcia-Burillo et al.(1998)]{gar98}
  Garcia-Burillo, S., Sempere, M.\ M., Combes, \& Neri, R.\
  1998, \aap, 333, 864
\bibitem[Gardiner \& Stone(2005)]{gar05}
  Gardiner, T.\ A., \& Stone, J.\ M.\ 2005, J.\ Comput.\ Phys., 205, 509
\bibitem[Hsieh et al.(2011)]{hsi11}
  Hsieh, P.-Y., Matsushita, S., Liu, G.\ Ho, P.\ T.\ P., Oi, N., \&
  Wu, Y.-L.\ 2011, \apj, 736, 129
\bibitem[Heller \& Shlosman(1994)]{hel94}
  Heller, C.\ H., \& Shlosman, I.\ 1994, \apj, 424, 84
\bibitem[Knapen et al.(2000)]{kna00}
  Knapen, J.\ H., Shlosman, I., \& Peletier, R.\ F.\ 2000, \apj, 529, 93
\bibitem[Knapen et al.(2002)]{kna02}
  Knapen, J.\ H., P\'erez-Ram\`irez, D., \& Laine, S.\ 2002, \mnras, 337, 808
\bibitem[Kim et al.(2008)]{kim08}
   Kim, C.-G., Kim, W.-T., \& Ostriker, E.\ C.\ 2008, \apj, 681, 1148
\bibitem[Kim \& Ostriker(2001)]{kim01}
   Kim, W.-T., \& Ostriker, E.\ C.\ 2001, \apj, 559, 70
\bibitem[Kim \& Ostriker(2006)]{kim06}
   Kim, W.-T., \& Ostriker, E.\ C.\ 2006, \apj, 646, 213
\bibitem[Kim et al.(2012)]{kim12}
   Kim, W.-T., Seo, W.-Y., Stone, J.\ M., Yoon, D., \& Teuben, P.\ J.\
   2012, \apj, 747, 60
\bibitem[Kormendy \& Kennicutt(2004)]{kor04}
   Kormendy, J., \& Kennicutt, R.\ C.\ 2004, \araa, 42, 603
\bibitem[Krause(1993)]{kra93}
   Krause, M.\ 1993, in IAU Symp.\ 157: The Cosmic Dynamo, eds.\ F.\ Krause,
   K.\ H.\ R\"adler, \& G.\ R\"udiger (Dordrecht:Kluwer), 305
\bibitem[Koldoba et al.(2002)]{kol02}
  Koldoba, A.\ V., Romanova, M.\ M., Ustyugova, G.\ V., \& Lovelace,
  R.\ V.\ E.\ 2002, \apj, 576, L53
\bibitem[Kulesza-\.Zydzik et al.(2009)]{kul09}
  Kulesza-\.Zydzik, B., Kulpa-Dybe\l, K., Otmianowska-Mazur, K.,  Kowal, G.,
  \& Soida, M.,\ 2009, \aap, 498, L21
\bibitem[Kulesza-\.Zydzik et al.(2010)]{kul10}
  Kulesza-\.Zydzik, B., Kulpa-Dybe\l, K., Otmianowska-Mazur, K., Soida, M.,
  \& Urbanik, M.\ 2010, \aap, 522, 61
\bibitem[Kulpa-Dybe\l\ et al.(2011)]{kul11}
  Kulpa-Dybe\l, K., Otmianowska-Mazur, K., Kulesza-\.Zydzik, B., Hanasz, M.,
  Kowal, G., W\'olta\'nski, D., \& Kowalik, K.\ 2011, \apj, 733, L18
\bibitem[Lou \& Fan(1998)]{lou98}
  Lou, Y.\ Q., \& Fan, Z.\ 1998, \apj, 493, 102
\bibitem[Maciejewski(2004)]{mac04}
  Maciejewski, W.\ 2004, \mnras, 354, 892
\bibitem[Maciejewski \& Sparke(2000)]{mac00}
  Maciejewski, W., \& Sparke, L.\ S.\ 2000, \mnras, 313, 745
\bibitem[Maciejewski et al.(2002)]{mac02}
  Maciejewski, W., Teuben, P.\ J., Sparke, L.\ S., \& Stone, J.\ M.\
  2002, \mnras, 329, 502
\bibitem[Maoz et al.(2001)]{mao01}
  Maoz, D., Barth, J., Ho, C., Sternberg, A., \& Filippenko, V.\
  2001, \aj, 121, 3048
\bibitem[Martini et al.(2003a)]{mar03a}
  Martini, P., Regan, M.\ R., Mulchaey, J.\ S., \& Pogge, R.\ W.\
  2003a, \apjs, 146, 353
\bibitem[Martini et al.(2003b)]{mar03b}
  Martini, P., Regan, M.\ R., Mulchaey, J.\ S., \& Pogge, R.\ W.\
  2003b, \apj, 589, 774
\bibitem[Martinez-Valpuesta et al.(2006)]{mar06}
  Martinez-Valpuesta, I., Shlosman, I., \& Heller, C.\ 2006, \apj, 637, 214
\bibitem[Mazzuca et al.(2008)]{maz08}
  Mazzuca, L.\ M., Knapen, J.\ H., Veilleux, S., \& Regan, M.\ W.\
  2008, \apj, 174, 337
\bibitem[Mazzuca et al.(2011)]{maz11}
  Mazzuca, L.\ M., Swaters, R.\ A., Knapen, J.\ H., \& Veilleux, S.\
  2011, \apj, 739, 104
\bibitem[Moss(1998)]{mos98}
   Moss, D.\ 1998, \mnras, 297, 860
\bibitem[Moss et al.(1998)]{mos_etal98}
   Moss, D., Korpi, M., Rautiainen, P., \& Salo, H.\ 1998, \aap, 329, 895
\bibitem[Moss et al.(1999)]{mos99}
   Moss, D., Rautiainen, P., \& Salo, H.\ 1999, \mnras, 303, 125
\bibitem[Moss et al.(2001)]{mos01}
   Moss, D., Shukurov, A., Sokoloff, D., Beck, R., \& Fletcher, A.\ 2001,
   \aap, 380, 55
\bibitem[Moss et al.(2007)]{mos07}
   Moss, D., Snodin, A., Englmaier, P.\ et al.\ 2007, \aap, 465, 157
\bibitem[Otmianowska-Mazur et al.(1997)]{otm97}
   Otmianowska-Mazur, K., von Linden, S., Lesch, H., \& Skupniewicz, G.\ 1997,
   \aap, 323, 56
\bibitem[Otmianowska-Mazur et al.(2002)]{otm02}
   Otmianowska-Mazur, K., Elstner, D., Soida, M., \& Urbanik, M.\ 2002, \aap,
   384, 48
\bibitem[Parker(1971)]{par71}
   Parker, E.\ N.\ 1971, \apj, 163, 255
\bibitem[Patsis \& Athanassoula(2000)]{pat00}
   Patsis, P.\ A., \& Athanassoula, E.\ 2000, \aap, 358, 45
\bibitem[Pease(1917)]{pea17}
   Pease, F.\ G.\ 1917, \apj, 46, 24
\bibitem[Piner et al.(1995)]{pin95}
  Piner, B.\ G., Stone, J.\ M., \& Teuben, P. J.\ 1995, \apj, 449, 508
\bibitem[Priest(1982)]{pri82}
  Priest, E.\ R.\ 1982, Solar Magnetohydrodynamics (Dordrecht: Holland), p199
\bibitem[Prieto et al.(2005)]{pri05}
  Prieto, M.\ A., Maciejewski, W., \& Reunanen, J.\ 2005, \aj, 130, 1472
\bibitem[Regan \& Mulchaey(1999)]{reg99}
  Regan, M.\ W., \& Mulchaey, J.\ S., 1999, \aj, 117, 2676
\bibitem[Regan \& Teuben(2003)]{reg03}
  Regan, M.\ W., \& Teuben, P.\ J.\ 2003, \apj, 582, 723
\bibitem[Regan \& Teuben(2004)]{reg04}
  Regan, M.\ W., \& Teuben, P.\ J.\ 2004, \apj, 600, 595
\bibitem[Regan \& Vogel(1995)]{reg95}
  Regan, M.\ W., \& Vogel, S.\ N.\ 1995, \apj, 452, L21
\bibitem[Rohde et al.(1999)]{roh99}
  Rohde, R., Beck, R., \& Elstner, D.\ 1999, \aap, 350, 423
\bibitem[Sandage(1961)]{san61}
  Sandage, A.\ 1961, The Hubble Atlas of Galaxies (Washington, DC:
  Carnegie Institution of Washington)
\bibitem[Sanders \& Huntley(1976)]{san76}
  Sanders, R.\ H., \& Huntley, J.\ M.\ 1976, \apj, 209, 53
\bibitem[Sanders \& Prendergast(1974)]{san74}
  Sanders, R.\ H., \& Prendergast, K.\ H.\ 1974, \apj, 188, 489
\bibitem[Shaw et al.(1993)]{sha93}
  Shaw, M.\ A., Combes, F., Axon, D.\ J., \& Wright, G.\ S.\ 1993, \aap,
  273, 31
\bibitem[Shaw et al.(1995)]{sha95}
  Shaw, M.\ A., Axon, D.\ J., Probst, R., \& Gatley, I.\ 1995, \mnras, 274, 369
\bibitem[Shlosman et al.(1989)]{shl89}
  Shlosman, I., Frank, J., \& Begelman, M.\ C.\ 1989, Nature, 338, 45
\bibitem[Shlosman et al.(1990)]{shl90}
  Shlosman, I., Begelman, M.\ C., \& Frank, J.\ 1990, Nature, 345, 679
\bibitem[Shukurov(1998)]{shu98}
  Shukurov, A.\ 1998, \mnras, 299, L21
\bibitem[Shukurov(2005)]{shu05}
  Shukurov, A.\ 2005, in Cosmic Magnetic Fields, ed.\ R.\ Wielebinski,
  \& R.\ Beck (Berlin: Springer), 129
\bibitem[Stone \& Gardiner(2009)]{sto09}
  Stone, J.\ M., \& Gardiner, T.\ A.\ 2009, NewA, 14, 139
\bibitem[Stone et al.(2008)]{sto08}
  Stone, J.\ M., Gardiner, T.\ A., Teuben, P.\, et al.\ 2008, \apjs, 178, 137
\bibitem[Thakur et al.(2009)]{tha09}
  Thakur, P., Ann, H.\ B., \& Jiang, I.\ 2009, \apj, 693, 586
\bibitem[van de Ven \& Fathi(2010)]{van10}
  van de Ven, G., \& Fathi, K.\ 2010, \apj, 723, 767
\bibitem[Wada \& Koda(2004)]{wad04}
  Wada, K., \& Koda, J.\ 2004, \mnras, 349, 270
\end{thebibliography}
\end{document}